\documentclass[11pt,preprint]{aastex}
\usepackage{graphics}
\usepackage{epsfig}
\usepackage{longtable}
\def\gsim{\lower 2pt \hbox{$\, \buildrel {\scriptstyle >}\over
{\scriptstyle \sim}\,$}}
\def\lsim{\lower 2pt \hbox{$\, \buildrel {\scriptstyle <}\over
{\scriptstyle \sim}\,$}}

\def\HI{H{\small I}}

\def\OVI{\ion{O}{6}}
\def\kms{km\,s$^{-1}$}

\shortauthors{}
\shorttitle{}

\begin{document}

\title{M31$^\ast$ and its circumnuclear environment}
\author{Zhiyuan Li\altaffilmark{1,2}, Q. Daniel Wang\altaffilmark{1}, Bart P. Wakker\altaffilmark{3}}
\altaffiltext{1}{Department of Astronomy, University of Massachusetts, 710 North Pleasant Street, Amherst, MA 01003, U.S.A.; zyli@head.cfa.harvard.edu, wqd@astro.umass.edu}
\altaffiltext{2}{Harvard-Smithsonian Center for Astrophysics, 60 Garden Street, MS-67, Cambridge, MA 02138, U.S.A.}
\altaffiltext{3}{Department of Astronomy, University of Wisconsin-Madison,
475 North Charter Street, Madison, WI 53705, U.S.A.; wakker@astro.wisc.edu}

\begin{abstract}

We present a multiwavelength investigation of the circumnuclear environment of
the Andromeda Galaxy (M31), utilizing archival {\sl Chandra}, {\sl FUSE}, {\sl GALEX, HST} and {\sl Spitzer} data as well as ground-based
observations. Based on the {\sl Chandra}/ACIS data, we tightly constrain the X-ray luminosity of M31$^\ast$, the central supermassive black hole (SMBH) of the galaxy, 
to be $L_{0.3-7 {\rm~keV}} \lesssim 1.2\times10^{36}{\rm~ergs~s^{-1}}$, approximately one part per 10$^{10}$ of the Eddington luminosity.  
From the diffuse X-ray emission, we characterize the circumnuclear hot gas with a temperature of $\sim$0.3 keV and a density of $\sim$$0.1{\rm~cm^{-3}}$. 
In the absence of an active SMBH and recent star formation, the most likely heating source for the hot gas is Type Ia supernovae (SNe). 
The presence of cooler, dusty gas residing in a nuclear spiral has long been known in terms of optical line emission and extinction. 
We further reveal the infrared emission of the nuclear spiral and evaluate the relative importance of various
possible ionizing sources. 
We show evidence for interaction between the nuclear spiral and the hot gas, 
probably via thermal evaporation. This mechanism lends natural understandings to 
1) the inactivity of M31$^\ast$, in spite of a probably continuous supply of gas from outer disk regions and 2) the launch of
a bulge outflow of hot gas, primarily mass-loaded from the circumnuclear regions.
One particular prediction of such a scenario is the presence of gas with intermediate temperatures arising from 
the conductive interfaces. The {\sl FUSE} observations do show strong \OVI$\lambda$1032 and $\lambda$1038 absorption lines against 
the bulge starlight, but the effective column density, 
$N_{\rm OVI} \sim 4\times10^{14}{\rm~cm^{-2}}$, may be attributed to foreground gas located in the bulge and/or
the highly inclined disk of M31, leaving the amount of circumnuclear gas with intermediate temperatures largely uncertain.
Our study strongly argues that stellar feedback, particularly in the form of energy release from SNe Ia, 
may play an important role in regulating the evolution of SMBHs and the interstellar medium (ISM) in galactic bulges.

\end{abstract}

\section{Introduction} {\label{sec:intro}}
Galactic circumnuclear environments, in which stars and the ISM are present in a dense state,  
are of vast astrophysical interest. 
The circumnuclear ISM, generally thought to be composed of externally acquired material and local stellar ejecta,
is the neccesary fuel for the SMBH to become an active galactic nucleus (AGN). Theoretically, how 
the fuel is transported to the SMBH is not fully understood (see a recent review by Wada 2004).
Observationally, direct links between nuclear activities and the properties of the ISM remain to be uncovered. 
Timescales of both dynamical and thermal processes are relatively short in circumnuclear 
regions. Thus a passive accumulation of the multi-phase (e.g., neutral and ionized) ISM there would inevitably lead to an interaction among its various
components, and possibly to an enhanced nuclear and/or star-forming activity (e.g., Ho, Filippenko \& Sargent 1997; Sarzi et al.~2007). 
That different phases of the ISM often co-exist in galactic
circumnuclear regions, some showing organized morphologies (e.g., van Dokkum \& Franx 1995; Macchetto et al.~1996; Knapen et al.~2005), suggests that there
are certain mechanisms regulating the dynamics and energetics of the ISM, which remain to be understood.
Furthermore, the global evolution of the ISM, one of the fundamental issues on galaxy evolution, can not be fully assessed 
without understanding its role in circumnuclear regions.

Owing to its proximity ($d\sim$780 kpc; 1$^\prime$ = 0.23 kpc), the Andromeda Galaxy (M31)
provides an ideal testbed for studing a galactic circumnuclear environment. Indeed, the inner few hundred parsecs of M31
have received vast observational attention, from radio to X-ray, tracing
all phases of the ISM and various types of stars in the region. As we will further discuss in
later sections,
it is clear that the SMBH of M31 (i.e., M31$^\ast$) is currently inactive and thus brings minimal disturbance to its environment, in which there is also little indication for
recent star formation.
Without the confusion from such activities,
this environment offers a unique ``quiescent'' close-up of 
possible relationships among the various interstellar and stellar components 
as well as the nucleus. In this work we aim to probe and to understand such relationships.

Studying the ISM against a high stellar radiation background is by no means straightforward at any wavelength. In particular, it is practically difficult to isolate the X-ray emission of truly diffuse hot gas
from the collective emission of individually unresolved stars.
In a recent study of the M31 bulge with {\sl Chandra} observations (Li \& Wang 2007; hereafter LW07), 
we have managed to account for the collective X-ray emissivity of faint, unresovled stars (see also Revnivtsev et al.~2007; Bogd\'{a}n \& Gilfanov 2008) according to the near-infrared (NIR) K-band
light distribution, and thus revealed the presence of diffuse hot gas on kpc-scales (LW07; Fig.~\ref{fig:dif}a). This procedure in turn 
allows us to advance the high-resolution study of hot gas in the circumnuclear regions.

The rich observational knowledge in the literature on M31$^\ast$ and its surrounding matter (except for the hot gas yet to be studied) 
deserves a selected summary in \S~\ref{sec:obs}, which in turn serves as a necessary guide for subsequent analyses and discussions.
The preparation of multiwavelength data used in this work is brifely 
described in \S~\ref{sec:data}. We present our analyses in \S~{\ref{sec:ana}}, particularly including quantifications
of the X-ray emission from M31$^\ast$ and of the hot gas, an analysis of \OVI\ absorption lines against the bulge starlight, 
and an examination of the circumnuclear dust emission. We discuss the results
in \S~{\ref{sec:dis}} and summarize our study in \S~{\ref{sec:sum}}.

\section{Observational knowledge on M31$^\ast$ and its environment} 
\begin{figure*}[!htb]
  \centerline{
    \epsfig{figure=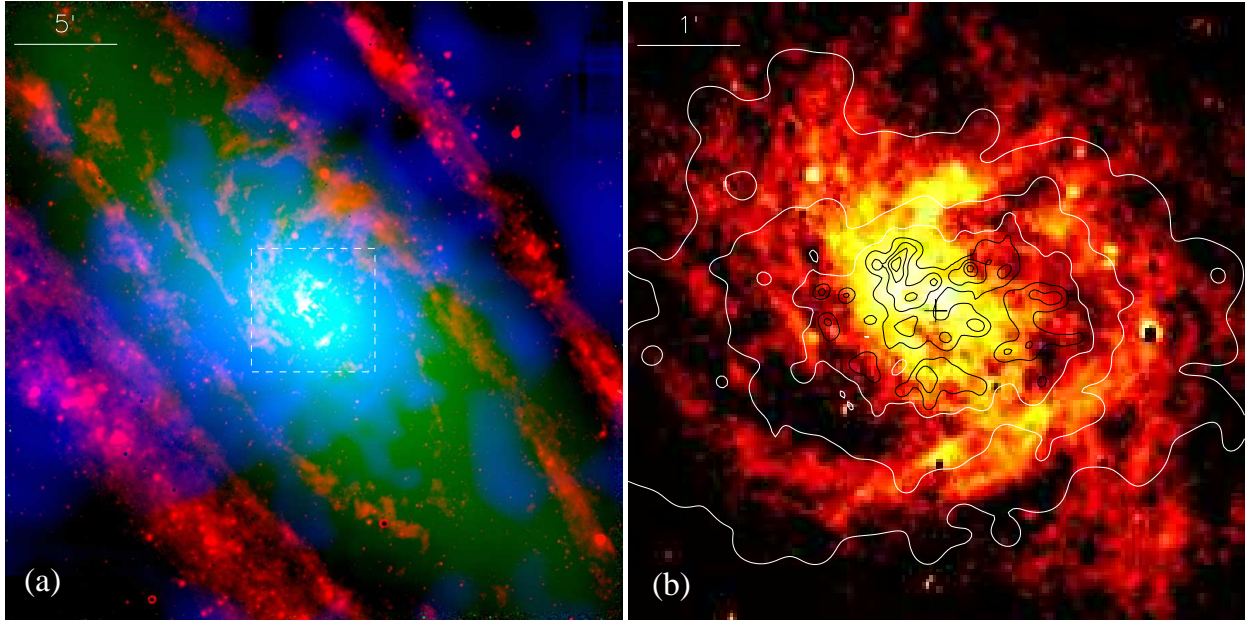,width=\textwidth,angle=0}
  }
  \caption{(a) Tri-color image of the central 30$^\prime$ by 30$^\prime$ (6.8 kpc by 6.8 kpc) of M31. {\sl Red}: {\sl Spitzer}/MIPS 24 $
\mu$m emission; {\sl Green}: 2MASS K-band emission; {\sl Blue}: {\sl Chandra}/ACIS 0.5-2 keV emission
of diffuse hot gas (LW07). The dashed box outlines the central 
6$^{\prime}$ by 6$^{\prime}$,  a region further shown in (b) and Fig~\ref{fig:ha_ir}. (b) Smoothed intensity contours of the 0.5-2 keV diffuse emission overlaid on the H${\alpha}$ emission. The contours are at 5, 10, 19, 27, 35, 45, and 55 ${\times}10^{-3}{\rm~cts~s^{-1}~arcmin^{-2}}$.
The plus sign marks the M31 center. }
 \label{fig:dif}
\end{figure*}

{\label{sec:obs}}
\begin{itemize}
\item {\bf Nucleus}

In the optical M31 hosts the well known double nuclei (so-called P1 and P2; Lauer et al.~1993) peaking at an angular separation of about half-arcsec from each other, 
which are interpreted to be an eccentric disk of typically K-type stars
with a total mass of $\sim$$2\times10^7 {\rm~M_{\odot}}$ (Tremaine 1995). P2 is fainter than P1 in V- and B-bands 
but brighter in U-band and in the UV, consistent with an addition of a 200 Myr-old starburst embedded in P2 
(called a third nucleus, P3; Bender et al.~2005). The SMBH is embedded in P2/P3, with an inferred dynamical mass of $\sim$$1.4\times10^8 {\rm~M_{\odot}}$ (Bender et al.~2005).

Crane, Dickel \& Cowan (1992) reported the possible detection of M31$^{\ast}$ in VLA 3.6 cm observations, giving a 
flux density of 28 $\mu$Jy.
Based on a 50 ks {\sl Chandra}/HRC observation, Garcia et al.~(2005; hereafter G05) claimed a 2.5 $\sigma$ detection of M31$^{\ast}$ with a 
0.3-7 keV intrinsic luminosity of $\sim9\times10^{35} {\rm~ergs~s^{-1}}$.

\item {\bf Stars}

The photometry in NIR (Beaton et al.~2007) and optical (Walterbos \& Kennicutt 1998) bands shows little color gradient
in the inner bulge. The colors are typical of an old, metal-rich stellar population, equivalent to type G5 III or K0 V. 
The Mg$_2$ index of 0.324 measured from the central $\sim$$30^{\prime\prime}$ (Burstein et al.~1988) indicates a metallicity
of [Fe/H]$\sim$0.3 (Buzzoni, Gariboldi \& Mantegazza 1992). For reference,
the K-band luminosity within the central 1$^\prime$ is 
4.7$\times10^{9} {\rm~L_{\odot,K}}$, which, according to the color-dependent (here a $B-V$ color of 0.95 is adopted) mass-to-light ratio of Bell \& de Jong (2001), corresponds
to a stellar mass of 4.0$\times10^{9} {\rm~M_{\odot}}$.

Using N-body simulations, Athanassoula \& Beaton (2006) showed that a classical bulge plus a bar-like structure is able to 
reproduce the observed NIR light distribution. This strongly argues for the presence of a stellar bar in M31, which is otherwise difficult to recognize
due to its high inclination. 

There is little evidence for any recent massive star formation in the 
circumnuclear regions. No massive (i.e., O and B-types) stars have been detected
(King et al.~1992; Brown et al.~1998); the far- to near-ultraviolet
($FUV-NUV$) color in the inner bulge suggests a stellar age over 300 Myr 
(Thilker et al.~2005); while a small amount of ionized gas is indeed present, it shows
optical line intensity ratios atypical of conventional HII regions (Rubin \& Ford 1971; del Burgo, 
Mediavilla \& Arribas 2000; see below).

\item {\bf Atomic gas}

So far there is no reported detection of atomic hydrogen in the central 500 pc; an upper lmit of $10^6 {\rm~M_{\odot}}$ is set on the HI mass (Brinks 1984).

\item {\bf Warm ionized gas}

The existence of ionized gas has long been known through the detection of [O II], [O III], H${\alpha}$, [N II] and [S II] emission lines in the spectra of the inner bulge (M${\rm \ddot{u}}$nch 1960; Rubin \& Ford 1971). Later narrow-band imaging observations (Jacoby, Ford \& Ciardullo 1985; Ciardullo et al.~1988; Devereux et al.~1994) further revealed that the gas is apparently located in a thin plane, showing filamental and spiral-like patterns, across the central few arcmins (so-called a {\sl nuclear spiral}; Fig.~\ref{fig:dif}b). 
The electron density of the ionized gas, inferred from the intensity ratio of [S II] lines, is $\sim$$10^2$-$10^4 {\rm~cm^{-3}}$ within the central arcmin, generally decreasing 
outward from the center (Ciardullo et al.~1988). The gas is estimated to have a mass of $\sim$$10^3{\rm~M_{\odot}}$, an H${\alpha}$+[N II] luminosity of a few $10^{39} {\rm~ergs~s^{-1}}$, and a very low volume filling factor consistent with its filamental morphology (Jacoby et al.~1985). 
The relatively high intensity ratio of [N II]/H${\alpha}$, ranging from $\sim$1.3-3 in different regions (Rubin \& Ford 1971; Ciardullo et al.~1988), is similar to the typical values
found in the bulge/halo of early-type galaxies (e.g., Macchetto et al.~1996) rather than in conventional HII regions (where typically [N II]/H${\alpha}$ $\sim$0.5).
The kinematics of the gas is rather complex. A major component of the velocity field apparently comes from
circular rotation, whereas the residuals indicate both radial and vertical motions (Rubin \& Ford 1971).  

That the stellar disk of M31 is probably barred offers a natural formation mechanism for the nuclear spiral: an inflow of gas from the outer disk 
driven by bar-induced gravitational purterbations to form organized patterns (e.g., Englmaier \& Shlosman 2000; Maciejewski 2004).
Indeed, by modelling the gas dynamics in a bar-induced potential Stark \& Binney (1994) obtained a satisfactory fit to the observed position-velocity diagram of the ionized and neutral gas in the central $\sim$2$^\prime$.
Another possible driver of gas is a recent head-on collision between M31 and its companion galaxy, most likely M32 (Block et al.~2006).  
Although details remain to be studied, it seems certain that an asymmetric gravitational potential is responsible for
the formation and maintenance of the nuclear spiral in M31, and perhaps so for similar gaseous structures found in the inner regions of disk galaxies (e.g., Regan \& Mulchaey 1999). 
The ionizing source of the nuclear spiral remains uncertain, however, especially in view of
the lack of massive stars {\sl in situ}. We discuss possible ionizing sources in \S~\ref{sec:dis}.

\item {\bf Dust and molecular gas}

Probing circumnuclear dust in M31 via optical extinction has a long history (e.g., Johnson \& Hanna 1972; Bacon et al.~1994; Sofue et al.~1994).
{\sl Spitzer} observations now provide the highest-resolution MIR/FIR view toward the circumnuclear regions. We show
below (\S~\ref{subsec:multi}) that these observations reveal the presence of interstellar dust and its remarkable association with the nuclear spiral, as pointed out by Sofue et al.~(1994) among others.

Detection of CO closest to the galactic center ($\sim$1\farcm3 away) points to a prominent dust complex, D395A/393/384, with an estimated molecular gas
mass of 1.5$\times10^{4} {\rm~M_{\odot}}$ (Melchior et al.~2000). 
This 100 pc-wide feature is also seen in the MIR/FIR emission (see \S~\ref{subsec:multi}).

\item {\bf Magnetic field and high energy particles}

At $10^{\prime\prime}$-$30^{\prime\prime}$ resolution the radio continuum emission shows filamentary patterns apparently associated with the H${\alpha}$ emission, i.e., the nuclear spiral (Walterbos \& Grave 1985; Hoernes, Beck \& Berkhuijsen 1998). 
The average power-law spectral index of $\alpha$$\sim$-0.75 ($S_\nu \propto \nu^\alpha$) throughout the 2.8-73.5 cm wavelength range indicates that the bulk emission is non-thermal (Walterbos \& Grave 1985).
Hoernes et al.~(1998) reported 
that the regular magnetic field appears to be oriented along the filaments. 
Assuming energy equipartition and a volume filling factor $f\sim1$, Hjellming \& Smarr (1982)
estimated an energy density of $\sim$$0.5{\rm~eV~cm^{-3}}$ for the energetic particles within the central 30$^{\prime\prime}$.

\end{itemize}

\section{Data preparation} {\label{sec:data}}
This work involves a variety of high-resolution data from IR to X-ray. Procedures of combining the archival {\sl Chandra}/ACIS-I
observations and spatially isolating the diffuse X-ray emission in the M31 bulge have been described in LW07. 
The same X-ray data set is used in this work to study the circumnuclear regions, for which an effective exposure of $\sim$90 ks 
is available. In order to maximize the counting statistics to constrain the X-ray emission of M31$^\ast$,   
we added to the data set a 38 ks ACIS-S observation (Obs.ID~1575; PI: S.~Murray). This observation unfortunately appears to be contaminated 
by a low-level, long-duration flare of cosmic-rays, hence we do not include it for the analysis of the diffuse emission, which requires a more stringent background filtering.
For the spectral analysis (\S~\ref{subsec:hotgas}), spectra extracted from individual observations were combined into a single spectrum.    
The considered energy range is restricted to 0.5-4 keV to minimize contamination by residual flares.
A ``stowed background'' was subtracted to remove the quiescent instrumental signals.
Although we do not have a precise knowledge of the sky background (LW07), its contribution to the analyzed spectra is negligible.

It is known that a bright X-ray source is located at $\sim$0\farcs5 from the position of M31$^\ast$ (G05; see \S~\ref{subsec:nuc}). Thus it is
a challenge for the ACIS image, with its 0\farcs49 pixel size and typical PSF FWHM of 0\farcs6-0\farcs7 near the optical axis, to isolate the emission from M31$^\ast$. 
Therefore we have followed the ``sub-pixel event repositioning'' technique
(Tsunemi et al.~2001; Li et al.~2003) to take advantage of sub-pixel information
from the dithering of the ACIS observations. The resultant ``super-resolution'' ACIS image has a PSF FWHM of $\sim$0\farcs5.
 
To trace the warm ionized gas we rely on the H$\alpha$+[N II] image of Devereux et al.~(1994) with $\sim$$2^{\prime\prime}$ resolution. 
For simplicity, in the following we refer to this image as the H$\alpha$ emission unless the [N II] component is explicitly specified.
We also obtained the {\sl Spitzer}/IRAC (Program ID 99; PI: M.~Rich), {\sl Spitzer}/MIPS (Program ID 3400; PI: G.~Rieke), 2MASS K-band (Jarrett et al.~2003), 
{\sl HST}/ACS F330W (Propos.ID 10571; PI: T.~Lauer) and {\sl GALEX} NUV/FUV (Gil de Paz et al.~2007) 
images of M31 from public archives, in order to provide a multiwavelength, co-spatial view of the various stellar and 
interstellar components. 
In particular, the {\sl Spitzer} MIR/FIR images offer unprecedented information on how the interstellar dust, and thus the cold neutral gas to some extent, is distributed across the circumnuclear 
regions. We note that, while the bulk of the FIR emission presumably arises from interstellar dust, a substantial fraction of the MIR emission from the inner bulge comes
from stellar objects, i.e., emission of the 
circumstellar envelopes around asymptotic giant branch (AGB) stars (e.g., Bressan, Granato \& Silva 1998), the spatial distribution of 
which closely follows the bulge starlight (e.g., Barmby et al.~2006; Gordon et al.~2006). Therefore we have subtracted a normalized K-band image from the IRAC 8 $\mu$m and MIPS 
24 $\mu$m images, respectively, to remove
the stellar contribution. To do so, we first constructed the radial (8 ${\mu}m-K$) and (24 ${\mu}m-K$) color profiles from consecutive annuli. Within each annulus the intensity of each band was chosen
to be the median (instead of azimuthally-averaged) value to minimize fluctuation introduced by
the interstellar component. Both profiles show little ($\lesssim$10\%) radial variation within the central $\sim$$3^\prime$ and thus represent the (8 ${\mu}m-K$) and (24 ${\mu}m-K$) flux ratios of the stellar component, which were then adopted as 
normalization factors for the K-band image.   
The $\lesssim$10\% uncertainty indroduced by the subtraction is expected to have little effect on the interpretation of subsequent analysis.
Hereafter the 8 and 24 $\mu$m emission refer to the interstellar component only.  


To probe the presence of gas with 
intermediate temperatures (a few $10^5$ K), we utilized {\sl FUSE} spectroscopic observations toward the center of M31,  
which is part of a program (Program ID C128; PI T.~Brown) aimed at understanding the stellar populations in the
cores of elliptical galaxies. Four exposures were taken covering the central 30$^{{\prime}{\prime}}$ by 30$^{{\prime}{\prime}}$, with a total useful exposure of 49~ks. 
The reduction and calibration of the {\sl FUSE} spectra were described in detail
by Wakker et al.\ (2003) and Wakker (2006). 

\section{Analysis and results} {\label{sec:ana}}
Our mining of the multiwavelength data is presented in this section. First we attempt to constrain
the amount of X-ray emission from M31$^\ast$, with much improved counting statistics compared to that achieved by G05.
Next we study the physical properties of the diffuse hot gas that, as we shall show below, fills the bulk of the circumnuclear volume. 
This volume-filling gas may naturally play a crucial role in 
regulating the mass and energy flows in the region.
From the multiwavelength view we then seek clues about physical relations
among the various ISM components.

\subsection{X-rays from M31$^\ast$} 
\begin{figure}[!htb]
 \centerline{
       \epsfig{figure=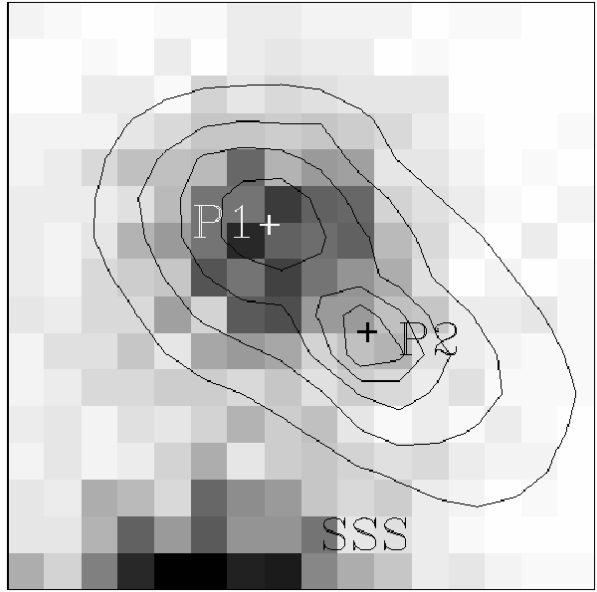,width=0.5\textwidth,angle=0}
 }
 \caption{A super-resolution 0.5-8 keV ACIS counts image (\S~\ref{sec:data}) in 0\farcs125 pixels, with the {\sl HST}/ACS F330W intensity contours
showing the double nuclei P1 and P2. The greyscale linearly ranges from 0 to 40 cts/pixel. The `+' signs mark the fitted centroids of P1 and P2. 
The displacement between P1 and P2 in X-ray is assumed to be same as in the optical. Part of the SSS appears at the bottom of the field.}
\label{fig:nuc}
\end{figure}

{\label{subsec:nuc}}
Fig.~\ref{fig:nuc} shows the distribution of 0.5-8 keV ACIS counts detected from the central $3^{\prime\prime} \times 3^{\prime\prime}$ region, along with an 
{\sl HST}/ACS F330W image showing the double nuclei P1 and P2. Dominating the X-ray emission are the two known bright sources with luminosities of $\sim$$10^{37} {\rm~ergs~s^{-1}}$: 
a super-soft source (SSS; partly shown in Fig.~\ref{fig:nuc}) is located
at $\sim2^{\prime\prime}$ south to the nuclei; the other source (named N1 by G05) is positionally coincident with P1, the presence 
of which severely hampers the isolation of the X-ray emission from M31$^\ast$.   

We assume that the X-ray source N1 is indeed embedded in P1. By fixing the angular displacement between N1 and the yet unresolved M31$^\ast$ as
that between the two optical nuclei, i.e., 0\farcs5 (Lauer et al.~1998, Fig.~8 therein), we perform a 2-d fit to the ACIS image by adopting a model 
consisting of three delta functions convolved with the local PSF, respectively respresenting SSS, N1 and M31$^\ast$, and a constant representing
the background ``diffuse'' emission. The local PSF is mimicked by stacking seven bright sources detected within the central 30$^{{\prime}{\prime}}$. The centroids of SSS and
N1, the amplitudes of all three sources and the background constant are determined by the fit, while the relative location of N1 and M31$^\ast$ is fixed
as assumed. The best-fit is achieved with an amplitude of 210$\pm$50 (1 $\sigma$) cts for M31$^\ast$. 
Corrected for the difference of effective area between ACIS-S3 and ACIS-I, the best-fit infers a 0.5-8 keV ACIS-I count rate of 1.4$\times10^{-3} {\rm~cts~s^{-1}}$,
corresponding to a 0.3-7 keV intrinsic luminosity of 1.2$\times10^{36} {\rm~ergs~s^{-1}}$
(assuming an absorbed power-law spectrum with $N_{\rm H}$$\sim$$7\times10^{20} {\rm~cm^{-2}}$, the Galactic foreground absorption column, and a photon-index of 1.7), which is
consistent with the 2.5 $\sigma$ detection of 9$\times10^{35} {\rm~ergs~s^{-1}}$ for M31$^\ast$ reported by G05 (based on $\sim$10 {\sl Chandra}/HRC  counts).
We have also tried to probe flux variation by examing the detected count rate of the 0\farcs25 by 0\farcs25 square enclosing M31$^\ast$ from
individual observations of typically 5 ks long. We find no statistically significant variation over a factor of 2 
from observation to observation as well as from within the 38 ks ACIS-S observation.

Our constraint for the X-ray emission of M31$^\ast$ relies on the assumption
that the bright X-ray source N1 is embedded in P1 instead of being an interloper. 
In galactic bulges the most likely interlopers are low-mass X-ray binaries (LMXBs). 
Gilfanov (2004) derived an empirical relation between the number of LMXBs with luminosites over $10^{37}{\rm~ergs~s^{-1}}$ and 
the stellar mass of the host galaxies, being $N_{\rm LMXB} \approx 14 {\rm~sources~per~10^{10}~M_\odot}$.  
Along the line of sight through the central arcsec, the observed K-band light  
infers a stellar mass of few $10^7{\rm~M_\odot}$. Accordingly, the expected number of LMXBs along this line is $\lesssim$0.05.
Hence the probability that we have detected N1 as an interloper is small.
Moreover, stellar mass BHs and neutron stars (NS) formed in the central cusp of the bulge are predicted
to sink into the close vicinity (few parsecs) of the SMBH, due to dynamical friction on less massive background stars (Morris 1993). 
Such migrators thus have a chance to 
compete with the SMBH on the consumption of surrounding gas to become bright X-ray sources.
Nayakshin \& Sunyaev (2007) modeled such a process and found that the collective X-ray luminosity
of the migrated compact objects can be higher than that of the SMBH by up to two orders of magnitude, dependent
on the mass of the common accretion disk. 
This scenario lends further support to the above procedure of pinning N1 on the position of P1,
in a sense that N1 is the appearance of one or a few accreting stellar mass BHs, being about ten times brighter
than M31$^\ast$. In regard of the common origin of P1 and P2 (i.e., an eccentric stellar disk), it is possible
that the X-ray signals detected from the position of P2, which we have considered as originated from M31$^\ast$,
also arise from stellar mass BHs. The marginal flux variation detected does not offer much help in distinguishing
between a stellar mass BH origin and a SMBH origin.
It is worth to note that, at the position of P1, which is the apocenter of the eccentric
orbits, BHs (and stars) have small velocities and hence a greater chance to capture the surrounding gas 
at large accretion rates. The situation is likely opposite at the position of P2. 
In conclusion, the assumed source positions are physically plausible. 

\subsection{Diffuse hot gas} {\label{subsec:hotgas}}
Fig.~\ref{fig:dif}a shows the large-scale 0.5-2 keV diffuse X-ray emission in the M31 bulge (LW07). The emission, showing an elongated 
morphology along the minor-axis and shadows cast by the outermost spiral arm and the prominent star-forming ring (e.g., Gordon et al.~2006) on
the near (northwestern) side of the tilted disk, has led LW07 to suggest that the hot gas is in the form
of a bi-polar outflow. Fig.~\ref{fig:dif}b shows this emission in the circumnuclear regions, along with the
H${\alpha}$ emission tracing the nuclear spiral. 
As is the case on large-scales, the X-ray emission appears i) elongated approximately along the minor-axis, 
and ii) fainter at the northwestern side beyond the central arcmin, presumably due to absorption by some cold ISM located in front 
of the hot gas.
These suggest that the bi-polar outflow is launched primarily in the very inner bulge.

\begin{figure*}[!htb]
  \centerline{
          \epsfig{figure=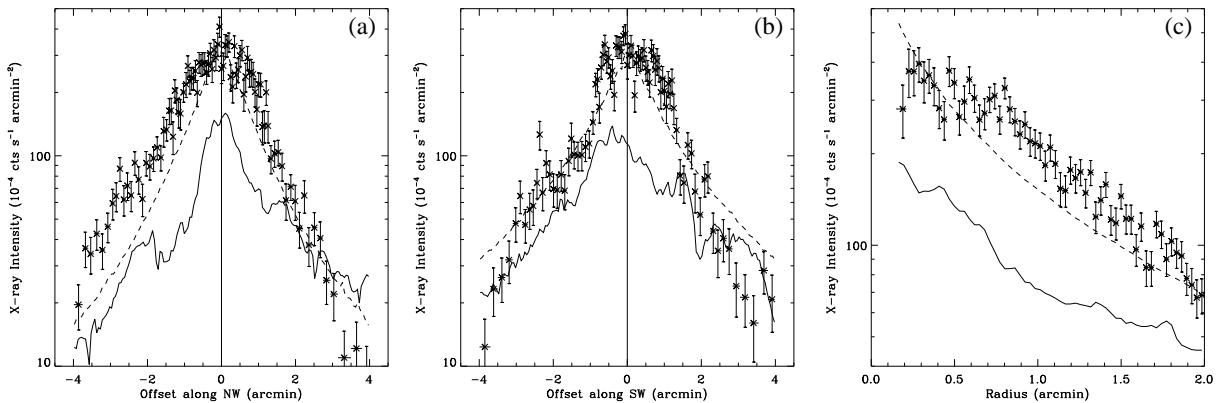,width=\textwidth,angle=0}
          }
	      \caption{0.5-2 keV diffuse X-ray intensity profiles along the minor-axis (a), the major-axis (b) and the radius (c). The vertical profiles are averaged within slices of 2$^\prime$ in width. 
A position angle of 40$^\circ$ is adopted.
In each panel, the dashed curve shows the corresponding K-band intensity profile with a 
normalization, $2.5\times10^{-5}{\rm cts~s^{-1}~arcmin^{-2}}/{\rm MJy~sr^{-1}}$, representing the {\sl already subtracted} contribution of unresolved X-ray sources (LW07), whereas the solid curve is
the corresponding H${\alpha}$ intensity profile with an arbitrary normalization. 
}
\label{fig:surbv}
\end{figure*}

A more quantitative view of the diffuse X-ray emission is shown in Fig.~\ref{fig:surbv} for the 0.5-2 keV intensity profiles 
along the minor (NW) and major (SW) axes and the radius. The intensity i) has a slightly steeper drop along the major-axis than along the minor-axis, and ii) is lower at the 
northwestern side along the minor-axis; these are fully consistent with the morphology shown in Fig.~\ref{fig:dif}b. 
The major-axis profile has a ``cap'' shape in the central $1^{\prime}$, where the intensity 
is more than two times higher than in the adjacent regions. Accordingly, the radial profile exhibits a flattening towards the center (Fig.~\ref{fig:surbv}c).
These apparently result from the clumpy X-ray emission positionally coincident with the nuclear spiral in the central arcmin (Fig.~\ref{fig:dif}b),  
indicating a relation between the two ISM components, although the X-ray intensity profile does not show a strict
correlation with the H${\alpha}$ intensity profile (solid curve in Fig.~\ref{fig:surbv}b).  
Compared to the steep decline of the K-band light (dashed curve in Fig.~\ref{fig:surbv}b), the ``cap''
also indicates a source of hot gas in addition to the stellar ejecta. 
A more detailed comparison of the multiwavelength emission will be given in \S~\ref{subsec:multi}.

\begin{figure*}[!htb]
  \centerline{
          \epsfig{figure=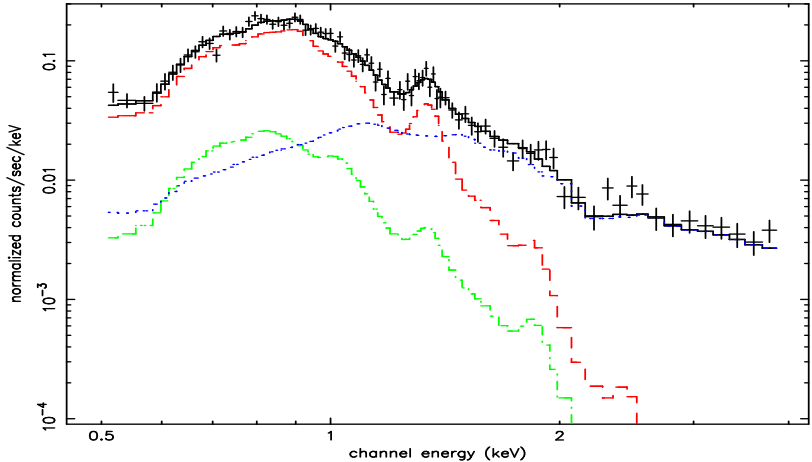,width=0.65\textwidth,angle=0}
          }
	      \caption{Spectrum of the central 1$^\prime$, fitted by a three-component model: VMEKAL
(gas; {\sl red curve}) + MEKAL (ABs; {\sl green curve}) + MEKAL (CVs; {\sl blue curve}). See text for details.}
\label{fig:spec}
\end{figure*}

To quantify the properties of the hot gas, we extract a representative spectrum of the total unresolved X-ray emission (i.e., from both the gas and unresolved stellar objects) 
from the central 1$^\prime$ (Fig.~\ref{fig:spec}). 
The unresolved stellar objects, with individual luminosities below 10$^{35}{\rm~ergs~s^{-1}}$, are predominantly cataclysmic variables
and coronally active binaries (CVs and ABs; Sazonov et al.~2006), the average spectrum of which unfortunately can not be determined in the M31 bulge 
due to the presence of hot gas. Instead, we rely on the spectral information derived from the dwarf elliptical galaxy, M32, which essentially 
lacks diffuse hot gas (Revnivtsev et al.~2007). In agreement with Revnivtsev et al.~(2007), we find that the spectrum of the unresolved X-ray emission from M32 can be characterized 
by a combination of two thermal plasma emission components (MEKAL in XSPEC), with temperatures of $\sim$0.4 and $\sim$4.6 keV and a solar abundance. 
The low- and high-temperature components are dominated by the emission of ABs and CVs, respectively. 
We adopt these two components to account for the stellar contribution in the M31 spectrum, fixing the temperatures and abundance and having the normalizations
scaled according to the underlying K-band light. 
As shown in Fig.~\ref{fig:spec}, the adopted model accounts well for the observed spectrum at energies above $\sim$1.5 keV. 
At lower energies, there is an indication that the stellar emissivity inferred for the M31 bulge is $\sim$1.5 times higher than that for M32 (LW07). 
We note that neglecting this difference has little effect on subsequent fit results, as the gas component dominates the 0.5-1.5 keV emission.

We then introduce a third thermal component (VMEKAL in XSPEC) 
to characterize the emission of hot gas, allowing the abundance, if other than solar as required by the fit, to be different among heavy elements. 
This component is subject to possible absorption by the cold ISM in the circumnuclear regions (see \S~\ref{subsec:multi}).
The fit is initiated with the abundance of all elements fixed at solar, giving a fitted temperature of $\sim$0.3 keV. Such a fit is poor, however. In particular,
it fails to account for a prominent line feature present at $\sim$1.33 keV,
most likely due to Mg XI K$\alpha$, given a gas temperature of $\sim$0.3 keV.
Therefore, we let the abundance of Mg be determined in a fit.
The abundance of Fe is also allowed to vary in the fit to account for  the Fe L-shell lines dominating energies of $\sim$0.8-1 keV.
The resultant fit, with $Z_{\rm Mg}$$\sim$2 and $Z_{\rm Fe}$$\sim$1, still shows
considerable discrepancies at energies below 0.7 keV, where lines of O VII 
and O VIII
dominate. Finally an acceptable fit is achieved with a fitted $Z_{\rm O}\sim0.4$.
The various fit results are summerized in Table~\ref{tab:spec}. We infer from the best-fit an electron density of 0.08${\rm~cm^{-3}}$
and a 0.5-8 keV unabsorbed flux of $1.0\times10^{-12}{\rm~ergs~cm^{-2}~s^{-1}}$ for the hot gas.

\begin{deluxetable}{ccccccccc}
\tabletypesize{\footnotesize}
\tablecaption{Fits to the spectrum of diffuse X-ray emission from the central arcmin}
\tablewidth{0pt}
\tablehead{
\colhead{Model} &
\colhead{$N_{\rm H}$$^a$} & 
\colhead{$T$} & 
\colhead{$\alpha$} &
\colhead{$Z_{\rm O}$$^b$} & 
\colhead{$Z_{\rm Fe}$$^b$} & 
\colhead{$Z_{\rm Mg}$$^b$} & 
\colhead{$EM$} &
\colhead{${\chi}^2/{\rm d.o.f.}$} \\
 & $10^{20}{\rm~cm^{-2}}$ & keV & & & & & ${\rm cm^{-6}~pc}$ &  
}
\startdata
VMEKAL &  $21^{+5}_{-8}$ & $0.28^{+0.03}_{-0.03}$ & - & 1  & 1 & 1 & $2.0^{+0.9}_{-0.8}$ &  112.5/85 \\
VMEKAL &  $19^{+8}_{-9}$ & $0.28^{+0.04}_{-0.04}$ & - & 1  & $1.1^{+0.3}_{-0.2}$ & $1.7^{+0.5}_{-0.5}$ & $1.8^{+1.2}_{-0.7}$ &  107.1/83 \\
VMEKAL &  $14^{+23}_{-14}$ & $0.25^{+0.06}_{-0.07}$ & - & $0.4^{+0.1}_{-0.2}$ & $0.9^{+1.0}_{-0.4}$ & $2.0^{+0.6}_{-0.5}$ & $2.8^{+3.0}_{-1.7}$ &  89.1/82 \\
CEVMAL &  15 & 3.3$^c$ & $-2.7^{+0.2}_{-0.2}$ & $0.2^{+0.2}_{-0.1}$ & $1.7^{+0.4}_{-0.3}$  & $2.2^{+0.7}_{-0.6}$ &  $40^{+39}_{-23}$  & 106.0/83
\enddata
\tablecomments{The quoted errors are at the  90\% confidence level. Values without errors are fixed in the fits. See text for details.
$^a$Including the Galactic foreground absorption. $^b$Abundance standard of Grevesse \& Sauval (1998) is applied. $^c$For $T_{\rm max}$.}
\label{tab:spec}
\end{deluxetable}  

The clumpiness of the circumnuclear X-ray emission (Fig.~\ref{fig:dif}b) implies that the hot gas is inhomogeneous, in particular, the gas 
could have more than one temperature. To try to account for this,
we fit the spectrum with the CEVMAL model in XSPEC. The model assumes a continuous distribution of emission measure ($EM$) such that 
$S_{\nu} = \int^{T_{\rm max}}_{T_{\rm min}}N\Lambda_\nu (T, Z) (T/T_{\rm max})^{\alpha} dT/T$ (Singh, White \& Drake~1996), where $S_{\nu}$ is the intrinsic spectrum,
$\Lambda_\nu$ the volume emissivity provided by the MEKAL model, and $N$ the normalization.
The maximum temperature $T_{\rm max}$ is fixed at 3.6 keV, the reason for which will become clear below (see \S~\ref{subsec:role}). Due to
the degeneracy between $T_{\rm max}$ and $\alpha$, the exact choice of $T_{\rm max}$ is not critical for the fit and its implications.
The minimum temperature $T_{\rm min}$ is effectively 0.08 keV in the implementation of MEKAL.
While the fit is not as satisfatory as the single-temperature fit, the resultant value of $\alpha$, being negative ($\sim$-2.7), implies that 
high-temperature gas, if it exists, only makes a minor contribution
to the total emission. If we further assume that the density and fractional volume of gas are power-law functions of temperature: 
$n \propto T^{\alpha_n}$, $dV/dT \propto T^{\alpha_V}$, from $S_{\nu} \equiv \int \Lambda_\nu (T, Z) n^2 dV$ we have
$\alpha=2{\alpha_n}+{\alpha_V}+1$. In the case of thermal pressure balance, ${\alpha_n} = -1$ and ${\alpha_V} = -1.7$, it can be shown 
that the volume-weighted temperature is $\sim$0.4 keV. This value is close to the single temperature of 0.3 keV derived from the spectral fit, i.e., emission weighted, suggesting that 
the diffuse X-ray emission arises from such a gas filling the bulk of the circumnuclear volume.

\subsection{Detection of \OVI\ absorption} {\label{subsec:ovi}}
Of course the above idea of the temperature distribution is just a simplification. That gas of lower temperatures is more 
emission measure-weighted, as implied by the negative value of $\alpha$, is not necessarily true but reflects the deficiency of our knowledge for 
emission at energies below $\sim$0.5 keV.   
Nevertheless, gas with intermediate temperatures, i.e., $\sim$10$^5-10^6$ K, may be traced by its emission or absorption against the background UV emission. It is known that the UV emission in the central few 
arcmins of M31 shows a bulge-like morphology and hence most likely arises from evolved stars (King et al.~1992; Thilker et al.~2005). 
In particular, the {\sl FUSE} spectrum (\S~\ref{sec:data}), covering $\sim$900-1200 \AA, is interpreted to be arising from 
hot subdwarf stars that are generally
thought to be responsible for the so-called {\sl UV upturn} (Brown 2004; see \S~\ref{subsec:uv}) typically observed in elliptical galaxies, 
as evidenced by the presence of the prominent photospheric Ly${\beta}$ and C III$\lambda$1175 absorption features (Brown et al.~1996).
On the other hand, \OVI\ $\lambda$1032 and $\lambda$1038 absorption lines, not characteristic of the photospheric spectral energy distribution (SED) of hot subdwarfs, are clearly
present, implying an interstellar origin.

\begin{figure*}[!htb]
  \centerline{
          \epsfig{figure=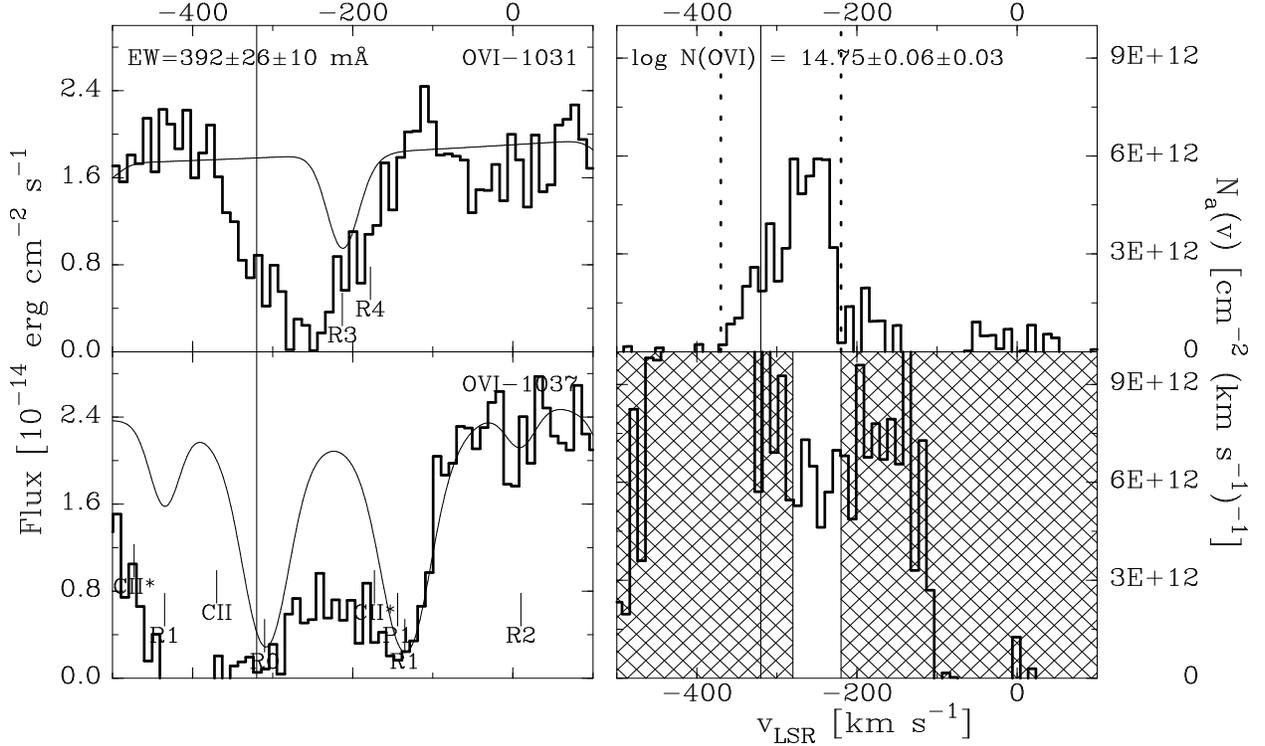,width=\textwidth,angle=0}
         }
	      \caption{{\sl FUSE} spectra of the OVI$\lambda\lambda$1031.926, 1037.617 absorption
lines. Left panels: flux; right panels: apparent column density.
Top panels: \OVI\ $\lambda$1031.926; bottom panels: \OVI\ $\lambda$1037.617. The solid histograms show the data, 
and the lines give
the continuum fit, which include a model for the H$_2$ absorption
lines. The R1, R2, R3, and R4 labels indicate the rest wavelengths
of H$_2$ lines. The vertical line gives the systemic velocity of M31.
In the right panels the dotted lines show the velocity range
over which the profile was integrated to derive the value of the \OVI\ column density that is shown. The cross-hatched region in the bottom
right panel shows the velocity range where the apparent column density
profile of the \OVI\ $\lambda$1037.617 line is contaminated by other
absorption lines.}
\label{fig:ovi}
\end{figure*}

We determine the continuum across the \OVI\ doublet by fitting a second-order polynomial
through absorption-line-free regions between 1028 and 1047~\AA. 
The resultant signal-to-noise ratio is 9.3 near the \OVI\ $\lambda$1032 line, with a flux of
1.9$\times10^{-14} {\rm~ergs~cm^{-2}~s^{-1}~\AA^{-1}}$. 
We also measure the H$_2$ column density, using the method described by Wakker (2006). H$_2$ is
detected both in the Milky Way and in M31. For the Milky Way we find
log\,$N$(H$_2$)=18.78, centered at a velocity of 1~\kms, while for M31 we find
log\,$N$(H$_2$)=15.19, centered at $-300{\rm~km~s^{-1}}$.
Fig.~\ref{fig:ovi} shows the resulting data (histogram) and continuum fit (continuous
line). The Galactic \OVI$\lambda$1032 absorption is relatively weak,
being log\,$N_{\rm OVI}=13.76\pm0.17\pm0.04$), as is the case for other sightlines in
this part of the sky (Wakker et al.\ 2003). Here the first error is the
statistical error associated with the noise in the data and the uncertainty in
the placement of the continuum, while the second error is the systematic error
associated with fixed-pattern noise and a 10~\kms\ uncertainty in the choice of
the velocity limits of the integration (cf.~Wakker et al.\ 2003 for details).

The \OVI$\lambda$1032 associated with M31 is contaminated by Galactic H$_2$
LP(3) 6-0 $\lambda$1031.191, while the \OVI$\lambda$1038 line is
contaminated by H$_2$ LR(1) 5-0 $\lambda$1037.149 and LR(0) 5-0
$\lambda$1036.545. After correcting for the H$_2$ line, we find an equivalent
width for the M31 \OVI$\lambda$1032 line of 392$\pm$26$\pm$10~m\AA. The
absorption profile can be converted into an apparent optical depth profile:
$N_a(v)=2.76\times10^{12}$ $log$(continuum/flux). By comparing the apparent optical
depth profiles of the two \OVI\ lines it is possible to assess whether the line
is saturated. We show the apparent optical depth profiles of the two lines in
the right two panels of Fig.~\ref{fig:ovi}, with the caveat that the \OVI$\lambda$1038 line can be
trusted only in a narrow range in velocities between $-$280 and
$-$220~\kms. In this velocity range, the apparent column densities of the two
\OVI\ lines match, showing that the \OVI$\lambda$1032 line is not saturated.
Then, integrating from $-$370 to $-$140~\kms, we find an effective \OVI\ column density of
log\,$N_{\rm OVI}=14.75\pm0.06\pm0.03$. Fitting the line with a Gaussian profile
gives a centroid velocity of $-$270~\kms and a Doppler parameter of 70~\kms.

It is not {\sl a priori} clear that all of this \OVI\ absorption is associated with
M31. In almost all sightlines in the region of sky within about
40$^\circ$ from M31 there are two high-negative velocity \OVI\ absorption
components, centered at about $-$300 and $-$150~\kms. Sembach et al.~(2003)
associated the former with a possible extension of the Magellanic Stream, while
the latter may represent gas distributed through Local Group, although this is a
tentative interpretation. Braun \& Thilker (2004) later discovered that there is
faint \HI\ emission in this part of the sky, with velocities that are continuous
with those of the Magellanic Stream. The \OVI\ column density in the
$-$300~\kms\ component is log\,$N_{\rm  OVI}=14.22\pm0.05$ in a sightline about
1\fdg9 from the M31 center (toward the QSO RX\,J0048.3+3941), which is at the
high end of \OVI\ column densities found for this component. Therefore, it is
likely that some fraction of the \OVI\ column density seen toward the M31 nucleus
originates in the Magellanic Stream and between the Milky Way and M31.
However, even if these contribute as much as log\,$N_{\rm OVI}=14.20$
(1.6$\times$10$^{14}$ cm$^{-2}$), the remainder (4$\times$10$^{14}$ cm$^{-2}$) is
likely associated with M31, implying the presence of gas with
temperatures of a few $10^5$ K. The next questions to address are how this gas is distributed within M31 and what 
is its possible origin. We shall discuss these issues in \S~\ref{sec:dis}.
 
\subsection{The circumnuclear regions in multiwavelength} 
\begin{figure*}[!htb]
\vskip -1cm
 \centerline{
  \epsfig{figure=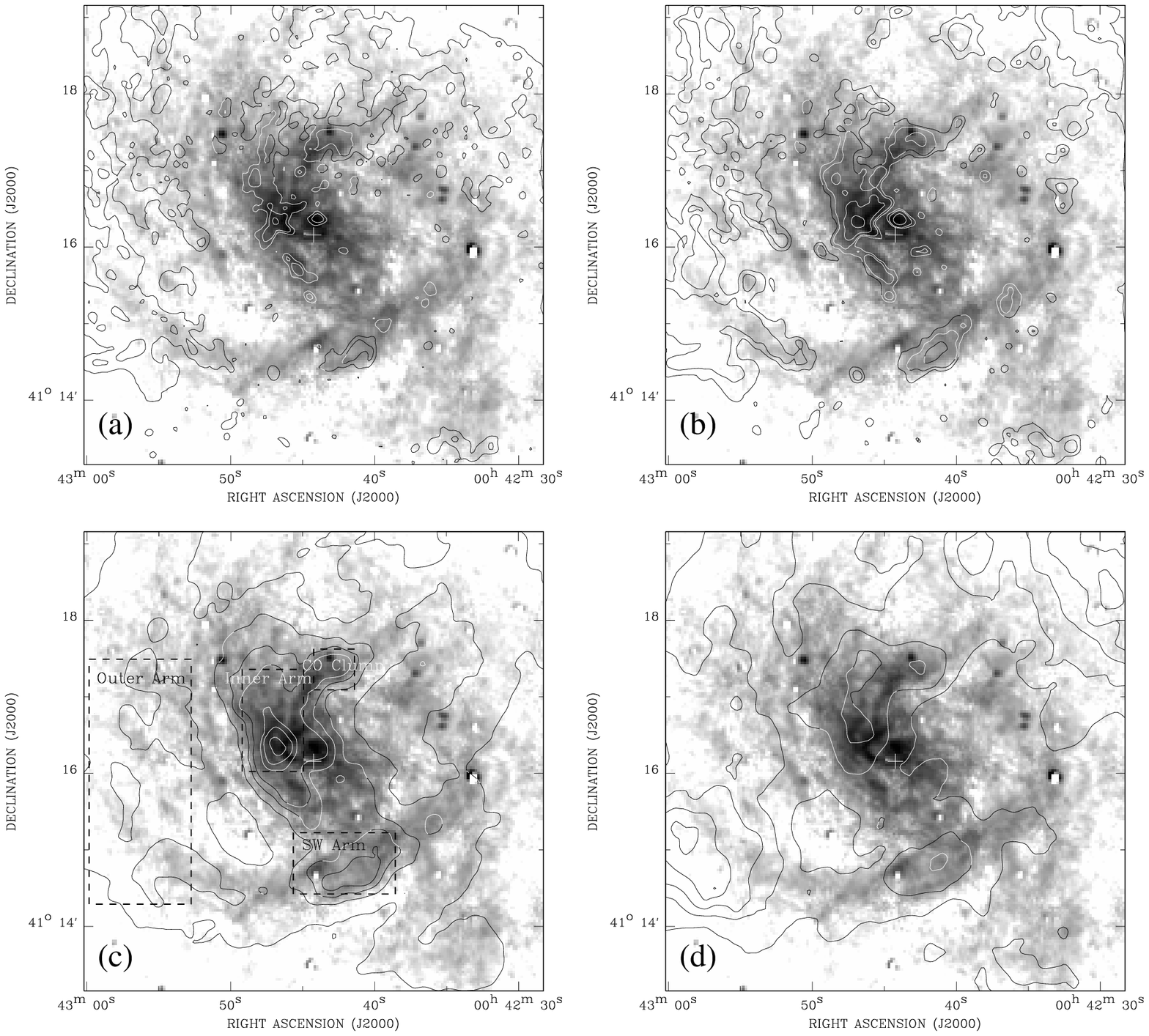,width=\textwidth,angle=0}
 }
    
   \caption{Contours of (a) 8 $\mu$m, (b) 24 $\mu$m, (c) 70 $\mu$m and (d) 160 $\mu$m emission overlaid on the H${\alpha}$ image 
of the central $6^\prime$ by $6^\prime$ region, in arbitrary units.
The dashed rectangles 
marked in (c) outline the selected regions for examination of multiwavelength correlations. See text for details.
   }
 \label{fig:ha_ir}   
\end{figure*}

{\label{subsec:multi}}
Fig.~\ref{fig:dif}a illustrates that structures of dusty gas are present in the inner disk regions.
The distributions of circumnuclear MIR and FIR emission are further shown in Fig.~\ref{fig:ha_ir}, along with the H$\alpha$ emission. 
A morphological similarity among the MIR, FIR and H$\alpha$ emission is evident, indicating that the interstellar dust is associated
with the ionized gas, i.e., they are both concentrated in the nuclear spiral. Apart from this overall similarity, region-to-region intensity contrasts among
the MIR, FIR and H$\alpha$ emission are also apparent. This is not unexpected, as both the strength of ionizing/heating 
sources and the density of the dusty gas could vary significantly across the circumnuclear regions. 

\begin{figure*}[!htb]
 \centerline{
  \epsfig{figure=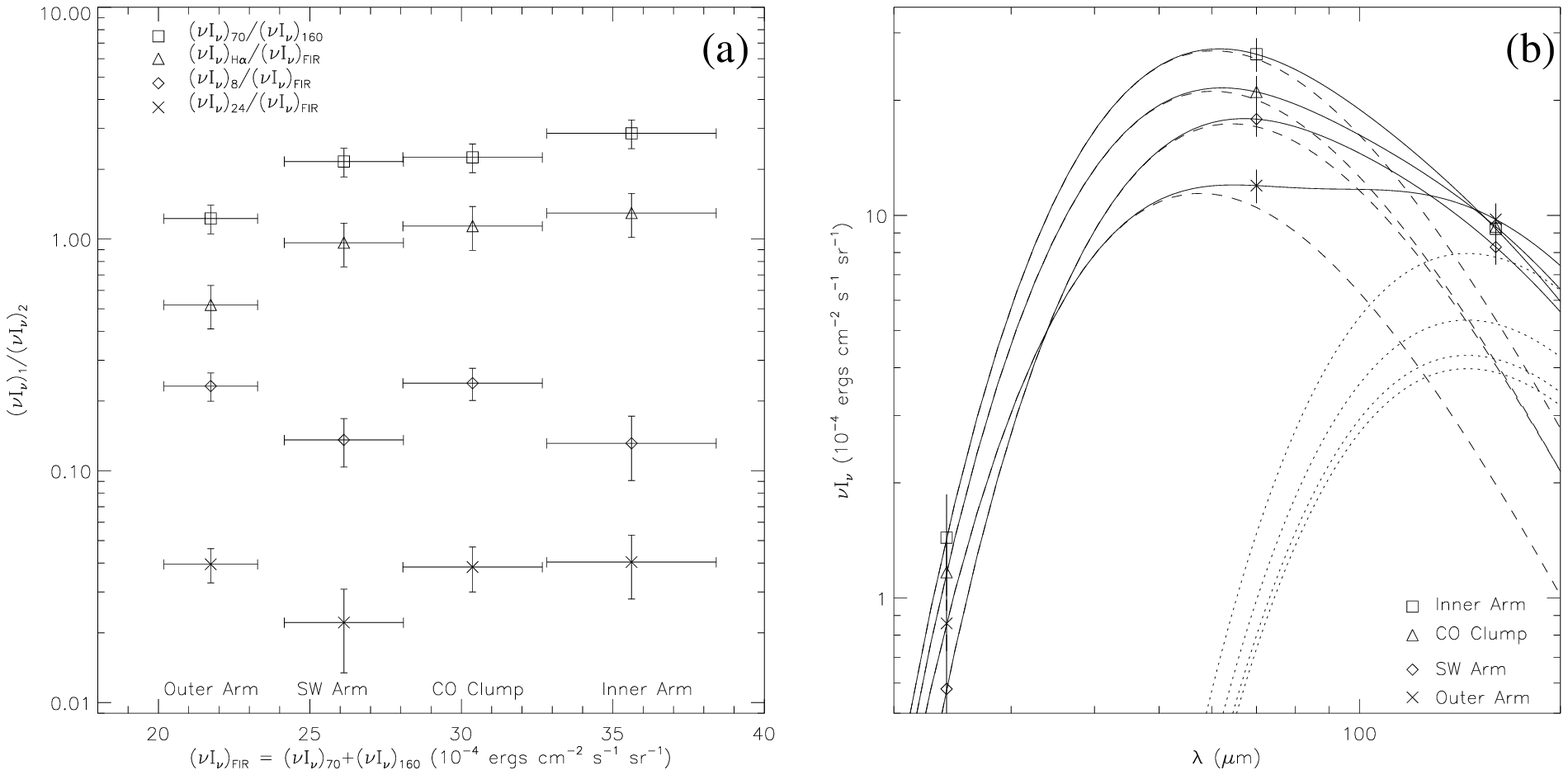,width=\textwidth,angle=0}
}
 \caption{(a) Intensity ratios as a function of the FIR intensity for selected circumnuclear regions. (b) The MIPS 24, 70 and 160 $\mu$m
intensities characterized by a two-component dust emission model (solid curves). The low- and high- temperature components are represented
by the dotted and dash curves, respectively. See text for details.}
\label{fig:rir}
\end{figure*}

We examine the multi-band intensity contrasts for several representative regions (Fig.~\ref{fig:ha_ir}c),
named as follows according to their positions and appearances. The ``Outer Arm'' is vertically present at $\sim$2\farcm5 east of the M31 center. Outward, it
spirals a winding angle of about $\pi$/2 and is rooted at the outermost spiral arm 
at northwest (Fig.~\ref{fig:dif}a);
inward, it turns northwestward and peaks in all the bands where it approximately intersects the major-axis at southwest,
and we refer to this region as the ``SW Arm''. The clump showing CO emission (\S~\ref{sec:obs}) is prominent in all the bands and is called the ``CO Clump''. Finally, an arcmin-long filament appears
coherent at northwest to the center and is referred to as the ``Inner Arm''. The multi-band intensities, defined as ${\nu}I_{\nu}$, of the four regions are measured from 
the images and summarized in Table~\ref{tab:ir}. We further show in Fig.~\ref{fig:rir}a for the individual regions various intensity ratios 
versus the FIR intensity, $({\nu}I_{\nu})_{\rm FIR}{\equiv}({\nu}I_{\nu})_{70}+({\nu}I_{\nu})_{160}$.
Although limited in number, the selected regions well sample the FIR intensity at values above $2.0\times10^{-3}{\rm~ergs~cm^{-2}~s^{-1}~sr^{-1}}$.
The ratio of $({\nu}I_{\nu})_{70}/({\nu}I_{\nu})_{160}$ increases with $({\nu}I_{\nu})_{\rm FIR}$, implying that
the stronger FIR emission is more weighted by warmer dust. Furthermore, both $({\nu}I_{\nu})_{70}/({\nu}I_{\nu})_{160}$ and
$({\nu}I_{\nu})_{\rm FIR}$ show the lowest (highest) value at the Outer (Inner) Arm, i.e., they generally 
increase with decreasing distances from the center. 
Such a trend is consistent with the bulge stellar radiation field, which is presumably responsible for heating the dust, being stronger at smaller galactocentric radii. 

\begin{deluxetable}{cccccccccc}
\tabletypesize{\scriptsize}
\tablecaption{Multiwavelength properties of the nuclear spiral in selected regions}
\tablewidth{0pt}
\tablehead{
\colhead{Region} &
\colhead{Position} &
\colhead{Size} &
\colhead{$({\nu}I_{\nu})_{\rm 8}$} &
\colhead{$({\nu}I_{\nu})_{\rm 24}$} &
\colhead{$({\nu}I_{\nu})_{\rm 70}$} &
\colhead{$({\nu}I_{\nu})_{\rm 160}$} &
\colhead{$({\nu}I_{\nu})_{\rm H_{\alpha}}$} & 
\colhead{$T_h$} &
\colhead{$M_d$} \\
(1) & (2) & (3) & (4) & (5) & (6) & (7) & (8) & (9) & (10)
}
\startdata
Outer Arm & (-136$^{\prime\prime}$, -16$^{\prime\prime}$) & 80$^{\prime\prime}$$\times$192$^{\prime\prime}$ & 5.0$\pm$0.6 & 0.9$\pm$0.1  & 12$\pm$1.2  & 9.8$\pm$1.0  & 11$\pm$2.3  & 50 & 28 \\
SW Arm & (24$^{\prime\prime}$, 80$^{\prime\prime}$) & 80$^{\prime\prime}$$\times$48$^{\prime\prime}$ & 3.6$\pm$0.8   & 0.6$\pm$0.2  & 18$\pm$1.8  & 8.3$\pm$0.8 & 25$\pm$5.0  & 45 & 3.9 \\
CO Clump & (16$^{\prime\prime}$, 72$^{\prime\prime}$) & 32$^{\prime\prime}$$\times$32$^{\prime\prime}$ & 7.2$\pm$1.0  & 1.2$\pm$0.2  & 21$\pm$2.1   & 9.3$\pm$0.9 & 35$\pm$6.9 & 48 & 1.3\\ 
Inner Arm & (-32$^{\prime\prime}$, 32$^{\prime\prime}$) & 48$^{\prime\prime}$$\times$80$^{\prime\prime}$ & 4.7$\pm$1.4 & 1.4$\pm$0.4  & 26$\pm$2.6 & 9.2$\pm$0.9  & 46$\pm$9.2 & 48 & 3.7
\enddata
\tabletypesize{\footnotesize}
\tablecomments{(1)-(3): Rectangular regions as defined in \S~\ref{subsec:multi}, their centroid positions with respect to the M31 center, and
sizes; (4)-(8): Multiwavelength intensities in units of $10^{-4}{\rm~ergs~cm^{-2}~s^{-1}~sr^{-1}}$. The quoted errors account for a 10\% calibration uncertainty 
for the MIR/FIR intensities and a 20\% for the H${\alpha}$ intensity. A 5-10\% uncertainty due to the removal of stellar contribution in the 8 and 24 $\mu$m emission 
is also propagated into the quoted errors; (9) Temperature of the warm dust component, in units of K; (10) Dust mass, in units of $10^3{\rm~M_{\odot}}$. }
\label{tab:ir}
\end{deluxetable}  

The ratio of $({\nu}I_{\nu})_{24}/({\nu}I_{\nu})_{\rm FIR}$, on the other hand, varies little among the different regions. It can be shown that
in each region the MIPS intensity ratios are inconsistent with a single dust temperature. A full accounting of the broadband IR SED 
requires a physical model of interstellar dust (e.g., Li \& Draine 2001), which is beyond the scope of the present study. Instead, we simply 
assume that the observed MIR/FIR emission can be reproduced by two distinct dust components, each characterized by 
the color temperature of a diluted blackbody, i.e., a $\lambda^{-1}$ emissivity law. Therefore in each region, the MIPS intensities
can be expressed by (e.g., Goudfrooij \& de Jong 1995):
\begin{equation}
   I_{\nu} = K_l \lambda^{-4}_{\mu} [e^{1.44{\times}10^4/({\lambda_{\mu}}T_l)}-1]^{-1} + K_h \lambda^{-4}_{\mu} [e^{1.44{\times}10^4/({\lambda{\mu}}T_h)}-1]^{-1},
{\label{eq:ir}}
\end{equation}
where $\lambda_{\mu}$ (= 24, 70, 160) and $I_\nu$ are in units of ${\mu}$m and ${\rm MJy~sr^{-1}}$, respectively. The 8 ${\mu}$m emission, presumably arising from line emission of PAH particles, 
is not considered here. Since the four unknowns in Eq.~\ref{eq:ir} can not be completely determined with 
the three MIPS bands, we further assume $T_l = 20 {\rm~K}$ for the cold dust component. The presence of cooler dust is possible, but its emission 
would be largely beyond the wavelength coverage of the MIPS. The remaining unknowns can then be solved for individual regions. The results (Table \ref{tab:ir}) indicate that 
the temperature of the warm dust component, $T_h$, falls in the range of 45-50 K, and that the relative contribution of this component to the
70 ${\mu}$m emission increases from 88\% at the Outer Arm to 97\% at the Inner Arm (Fig.~\ref{fig:ha_ir}b). We further estimate the dust mass according to 
$M_d = 5.1{\times}10^{-2}d^2(K_l+K_h) {\rm~M_\odot}$ (Goudfrooij \& de Jong 1995), where $d = 0.78$ is the distance of M31 in Mpc.
In particular, a dust mass of $1.3\times10^3 {\rm~M_\odot}$, or a gas mass of $1.3\times10^5 {\rm~M_\odot}$ (assuming a dust-to-gas mass ratio of 0.01), 
is inferred for the CO Clump. Corrected for the considered area, this value is about four times higher than
that derived from optical extinction (Melchior et al.~2000). 
Such a discrepancy, often encountered in the determination of dust mass in elliptical galaxies (e.g., Goudfrooij \& de Jong 1995), does not necessarily cast serious doubt on our results, 
in view of the disadvantage of an extinction study to probe a diffusely distributed component of dust. 
On the other hand, it is worth noting that the estimated mass is largely weighted by the low-temperature component and thus dependent on the less certain 
value of $T_l$. An adoption of $T_l = 25 {\rm~K}$, for instance, would result in typically 50\% less mass. The equivalent
hydrogen column density is inferred to be (1-2)${\times}10^{21}{\rm~cm^{-2}}$ for the four regions, consistent with the value inferred from
the X-ray spectral fit (Table \ref{tab:spec}).

\begin{figure*}[!htb]
 \epsfig{figure=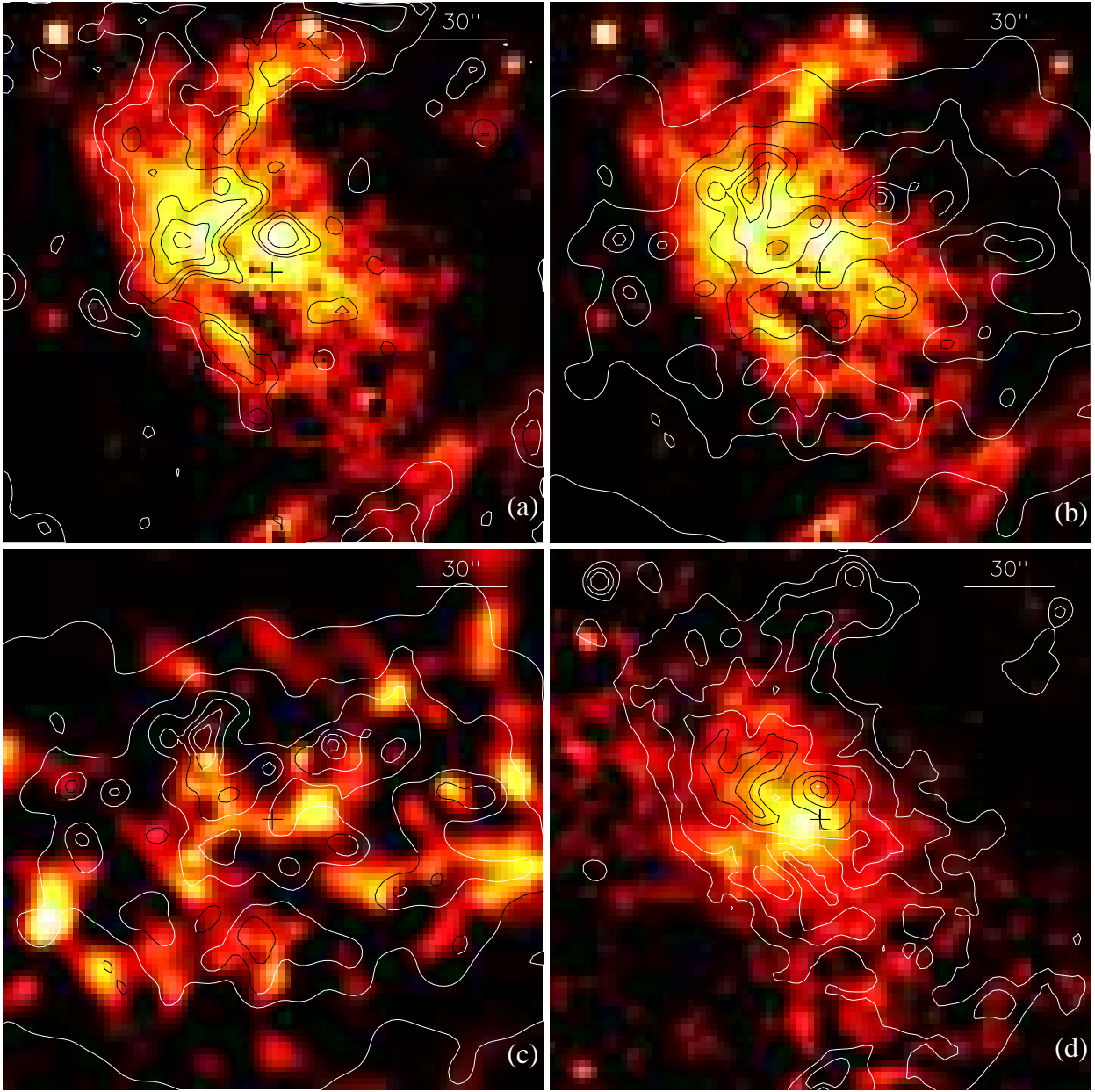,width=\textwidth,angle=0}
 \caption{A multiwavelength view of the central 3$^\prime$ by 3$^\prime$ region. (a) Contours of 24 $\mu$m emission overlaid on
the H${\alpha}$ image. (b) Contours of 0.5-2 keV diffuse X-ray emission overlaid on the H${\alpha}$ image. (c) 
Contours of 0.5-2 keV diffuse X-ray emission overlaid on the X-ray hardness ratio map. (d) Contours of H${\alpha}$ emission
overlaid on the ($NUV-FUV$) color map.}
\label{fig:multi}
\end{figure*}

A zoom-in view of cross-correlations among the 24 $\mu$m, H$\alpha$, UV and X-ray emission 
in the central $3^\prime$ by $3^\prime$ region is presented in Fig.~\ref{fig:multi}.
The FIR emission, limited in spatial resolution in this region, is expected to be represented by the 24 $\mu$m emission. 
Within the central $\sim$$1^\prime$, both the H$\alpha$ and 24 $\mu$m emission appear filamentary at the resolution of few arcseconds (Fig.~\ref{fig:multi}a).
It is now clear that the CO Clump is part of an arm-like filament. Inward, this filament 
joins the Inner Arm at $\sim$$30^{\prime\prime}$ east of the center, where the 24 $\mu$m emission peaks, and further extends southwestward
across the minor-axis. The association between the H$\alpha$ and 24 $\mu$m emission 
along these features is evident. There is no further trace of coherent 24 $\mu$m emission, however, on the southwestern side of the 
center, where the nuclear spiral is still prominent in H$\alpha$ emission.
Except for its presence at a bright knot
located $\sim$$15^{\prime\prime}$ north of the center,
the 24 $\mu$m emission is also largely absent towards the very central regions, 
consistent with the distribution of optical extinction (Bacon et al.~1994; Melchior et al.~2000).  
We note that this knot exhibits optical line emission as well, positionally coincident with ``cloud F'' designated by de Burgo et al.~(2000)
among their detection of several isolated features (clouds) of ionized gas within the central $\sim$$15^{\prime\prime}$. 
Interestingly, the H$\alpha$ emission, wherever it is seen associated
with the 24 $\mu$m emission, appears brighter on the side facing the galactic center. 

In the region near the nuclear spiral, the diffuse X-ray emission appears clumpy and enhanced with respect to the overall elongated morphology along the minor-axis (Fig.~\ref{fig:multi}b). 
The emission also shows a sign of flattening toward the galactic center, as implied in Fig.~\ref{fig:surbv}c.
As shown in Fig.~\ref{fig:multi}c, there is no clear correlation between the X-ray intensity and the hardness ratio,
defined as ($I_{1-2{\rm~keV}}-I_{0.5-1{\rm~keV}}$)/($I_{1-2 {\rm~keV}}+I_{0.5-1 {\rm~keV}}$), indicating that the X-ray clumpiness is not
merely a result of spatially varying obscuration. 
Rather, the observed X-ray enhancement close to the nuclear spiral, albeit a picture complicated by projection effects, suggests a physical relation between the hot and cooler gas. 

It is evident from the radial intensity distributions that the FUV emission 
increases more steeply toward the center than the NUV emission does (Thilker et al.~2005), 
suggesting a gradual change of the stellar UV SED. 
Such a trend is illustrated in Fig.~\ref{fig:multi}d, where the 2-D distribution 
of ($m_{\rm NUV}-m_{\rm FUV}$) is shown along with the H$\alpha$ emission. For reference, the color, $m_{\rm NUV}-m_{\rm FUV}$, has a mean value of 
$\sim$$-$0.95 within the central 30$^{\prime\prime}$, about 0.2 higher than that
in the regions immediately beyond. 
Most surprisingly, the overall morphology of the ($NUV-FUV$) color is similar to that of the nuclear spiral rather than
that of the bulge, indicating that the apparent rise of the FUV emission relative to the NUV emission is not simply 
due to a radial change in the stellar SED. 
In principle, the excess could be the result of differential extinction introduced 
by the cold ISM residing in the nuclear spiral. Quantitatively, adopting $A_{\rm FUV}/E(B-V) = 8.376$, $A_{\rm NUV}/E(B-V) = 8.741$ (Wyder et al.~2005;
based on the extinction law of Cardelli, Clayton \& Mathis 1989),  
$N_{\rm H}/E(B-V) = 6\times10^{21} {\rm~atom~cm^{-2}~mag^{-1}}$ and an equivalent hydrogen column of $10^{21}{\rm~cm^{-2}}$, 
the differential extinction is estimated to be $A_{\rm NUV}-A_{\rm FUV} \approx 0.06$ and not able to fully account for the observed color excess.
That there is no color excess seen at the position of the CO Clump
also indicates that differential extinction has a minor effect. 
Alternatively, the color excess could be due to intrinsic FUV emission associated with the nuclear spiral, a possibility to be further discussed below.

\section{Discussion} {\label{sec:dis}}
Based on the multiwavelength analyses above, we now explore the
physical nature of various phenomena and processes in the M31 
circumnuclear regions.
In the central few hundred parsecs, the ISM consists of two dynamically distinct components. One is the nuclear spiral with a low volume filling factor,
consisting of cold dusty gas, traced by the MIR and FIR emission, and warm ionized gas, traced by optical recombination lines.
The nuclear spiral is thought to be formed by bar-induced gravitational perturbations with a possibly continuous supply of gas from the outer disk regions.
Connections between the nuclear spiral and the major spiral arms in the outer disk are evident in Fig.~\ref{fig:dif}a (see also Gordon et al.~2006).
The other component is a corona of volume-filling hot gas, traced by the diffuse X-ray emission. This hot corona
has a bi-polar extent of at least several kpc
away from the midplane (LW07). 
While young massive stars are essentially absent, embedded in the hot corona is an old stellar population with a total mass of $\sim$$10^{10}{\rm~M_\odot}$, which is primarily responsible for the gravitational potential and probably for the energetics of the ISM. 
Finally, there is the inactive SMBH possibly manifesting itself in radio and X-ray to date.
Both the circular speed ($v_c\sim270 {\rm~km~s^{-1}}$ at $r\approx230 {\rm~pc}$) and the 
sound speed of the hot gas ($c_s\sim280 {\rm~km~s^{-1}}$ at a temperature of 0.3 keV)
imply a relatively short dynamical timescale of $\sim$$10^6$ yr. 
Unless our multiwavelength view is a highly transient one, which is unlikely, there ought to be 
certain physical processes regulating the behavior of the multi-phase ISM as well as that of the SMBH. 
In the following discussion we aim primarily to propose a self-consistent scenario for this regulation. 

For ease of quantification, we adopt fiducial values of the hot and warm ionized gas 
in the central 1$^\prime$ ($\sim$230 pc) as: $n_h = 0.1 {\rm~cm^{-3}}$, $T_h = 4\times10^6 {\rm~K}$, 
$n_w = 200 {\rm~cm^{-3}}$, $T_w = 10^4 {\rm~K}$, respectively. $n_h$ and $T_h$ are inferred
from the X-ray spectral fit (\S~\ref{subsec:hotgas}), while $n_w$ is roughly the volume-weighted average value inferred from [S II] line ratios (Ciardullo et al.~1988), given the canonical value of $T_w$ for ionized gas.
For both phases the notation of density is for hydrogen and, as assumed for simplicity, equally for electrons.
The density and temperature of the cold gas are 
less certain, and we adopt $n_c = 10^4{\rm~cm^{-3}}$ and $T_c$ = 100 K so that $n_cT_c \sim 2n_hT_h$, the case of thermal pressure balance. 
The slightly higher pressure of the warm gas can be understood if this phase represents an interface between the hot and cold phases (see \S~\ref{subsec:role}).
Because of the high pressure environment,
the bulk of hydrogen in the cold phase is likely in molecular form (Wolfire et al. 1995), which is also suggested by the estimated mass ($\sim$$5\times10^{6}{\rm~M_\odot}$) of cold gas in the central 500 pc much
exceeding the upper limit of atomic hydrogen mass ($\sim$$10^{6}{\rm~M_\odot}$) set by 21 cm observations (Brinks 1984).  
Now, the mass of hot gas, $M_h \approx 1.2\times10^5{\rm~M_\odot}$, is a straightforward measurement based on $n_h$ and the considered volume $V \approx 1.5\times10^{63}{\rm~cm^{3}}$.
The mass of warm gas, $M_w \approx 3.2\times10^3 {\rm~M_\odot}$, is estimated from the H$\alpha$ emission by assuming
a standard case B emissivity and a [N II]/H$\alpha$ intensity ratio of 2.
The mass of cold gas, $M_c \approx 10^6 {\rm~M_\odot}$, is inferred from the MIR/FIR emission using the procedure described in \S~\ref{subsec:multi}.

\subsection{Accretion and feedback of the nucleus} 
M31$^\ast$, with an 0.3-7 keV luminosity of 1.2$\times10^{36} {\rm~ergs~s^{-1}}$ if the observed emission
from P2 entirely arising from the SMBH (\S~\ref{subsec:nuc}), ranks as the second faintest galactic nucleus
detected in X-ray, after Sgr A$^\ast$, but comparable to M32$^\ast$ (Ho, Terashima \& Ulvestad 2003, who found
a 2-10 keV luminosity of $10^{36}{\rm~ergs~s^{-1}}$). Assuming that the X-ray 
emission typically accounts for $\sim$$10\%$
of the bolometric luminosity, we have $L_{\rm bol}\sim10^{37} {\rm~ergs~s^{-1}}$, in comparison to
the Eddington luminosity of $L_{\rm Edd} \approx 10^{46}(M_{\rm BH}/10^8{\rm~M_\odot}){\rm~ergs~s^{-1}}$,
where $M_{\rm BH}$ is the mass of the SMBH.
Either the hot gas
or the cold gas in the circumnuclear environment is sufficiently abundant to feed M31$^\ast$ for a time up to 10 Gyr at its present accretion rate.
For instance, assuming that M31$^\ast$ is powered by the Bondi accretion (1952)
of the soft X-ray-emitting hot gas, the accretion rate can be estimated as: 
\begin{eqnarray}
\dot{M}_{\rm Bondi} \approx 4{\pi}{\lambda}m_{\rm H}n_h(GM_{\rm BH})^2c_s^{-3}~~~~~~~~~~~~~~~~~~~~~~~~~~~~~~~~~~~~~~~~~~~~~~~ \\ \nonumber
\approx 5.5\times10^{-5}(\frac{n_h}{0.1{\rm~cm^{-3}}})(\frac{M_{\rm BH}}{10^8{\rm~M_\odot}})^2(\frac{c_s}{300{\rm~km~s^{-1}}})^{-3}{\rm~M_\odot~yr^{-1}},
\end{eqnarray} 
where ${\lambda}$ is a numerical factor taken to be 0.25 and the rest symbols are of conventional meanings if not yet defined above. 
The corresponding {\sl Bondi luminosity} is $L_{\rm Bondi} \equiv \eta\dot{M}_{\rm Bondi}c^2 \approx 3.1\times10^{41} (\eta/0.1){\rm~ergs~s^{-1}}$,
where $\eta$ is the radiation efficiency. It follows that $L_{\rm bol}/L_{\rm Bondi} \sim 10^{-5}$,
indicating that the radiation of M31$^\ast$ is highly inefficient, i.e., $\eta \ll 0.1$. 
The estimated Bondi accretion rate could be somewhat biased, as we do not have precise knowledge on the physical properties
of the accretion flow on pc-scales. Regardless, the radiation efficiency of M31$^\ast$ is likely low, 
as predicted by models of advection-dominated accretion flow with typically low accretion rates ($\dot{M}_{\rm acc}/\dot{M}_{\rm Edd} \ll 1$; Narayan \& Yi 1995).

On the other hand, it is not fully understood what mechanisms act to remove the angular momentum of gas originating 
from the $r$$\sim$100 pc regions and transport the gas to the sub-pc vicinity of the SMBH (Wada 2004). 
In any case, this theoretical difficulty does not seem to pertain to M31. We recall that 
there is no clear evidence of gas gathering into the central few parsecs of M31 (\S~\ref{subsec:multi});
any coherent entity of cold gas is apparently located at distances $\gtrsim$50 pc.
This is contradicted by the presence of gas inflow, favorably induced by the bar potential and
forming the nuclear spiral. It is sometimes suggested by numerical models (e.g., Maciejewski 2004) that 
the inflowing gas ultimately settles in semi-circular orbits, where the orbital energy is minimal for a given angular
momentum, forming a so-called {\sl nuclear ring}. Accumulations of gas on such orbits are expected to be subject to 
gravitational instabilities that leads to star formation,
by which the ring manifests itself in observations (e.g., Sarzi et al.~2007), 
or to a further infall to the galactic center, activating the nucleus. {\sl Neither}
situation is observed in M31, whereas there is no reason to argue that the 
global gas inflow has been stopped.        
In this regard, the right question to ask for M31 seems to be: what mechanism prevents the gas from gathering into the 
central few parsecs? We shall further address this issue below.

Alternatively, M31$^\ast$ can obtain fuel from its immediate surroundings, i.e., the eccentric stellar disk
consisting of old stars with a total mass of $\sim$$2\times10^7{\rm~M_\odot}$ (Tremaine 1995).
It is expected that these stars lose mass via stellar winds at a rate of $\sim$$10^{-4}{\rm~M_\odot~yr^{-1}}$ (e.g., Ciotti et al.~1991),
a value comparable to the above estimated Bondi accretion rate.
Chang et al.~(2007) argued that the stellar winds are gravitationally bound and would eventually collapse owing to radiative cooling to form a 
thin gaseous disk orbiting around the SMBH on pc-scales, and that star formation can be triggered from the gaseous disk every 0.1-1 Gyr, 
consistent with the 200 Myr starburst proposed for P3 (\S~\ref{sec:obs}).
These authors noticed that, given the estimated mass of a few 10$^3{\rm~M_\odot}$ for P3, the star formation efficiency
is $\sim$$10\%$-$20\%$. The remaining gas could then be accreted by the SMBH, although details of such a process remain unclear. 

It is not trivial to quantify feedback from galactic nuclei, even for the inactive ones such as M31$^\ast$. 
Much of the feedback is expected to be carried out by jets of relativistic particles, especially for ``faint''
nuclei showing a low radiation efficiency. Allen et al.~(2006) found an empirical relation between accretion rate 
and jet power in X-ray-bright elliptical galaxies, quantitatively as: ${\rm log}(L_{\rm Bondi}/10^{43}{\rm~ergs~s^{-1}}) 
= 0.65 + 0.77 {\rm log}(L_{\rm jet}/10^{43}{\rm~ergs~s^{-1}})$. The Bondi accretion rate of their sample nuclei
ranges from $3.5\times10^{-4}{\rm~M_\odot~yr^{-1}}$ to $4.6\times10^{-2}{\rm~M_\odot~yr^{-1}}$.
It is not known as {\sl a priori} how the relation behaves at lower accretion rates. If it holds for M31$^\ast$, the jet power would then be $\sim 
1.6\times10^{40}{\rm~ergs~s^{-1}}$, or about $5\%$ of the Bondi luminosity. In principle, this powerful nuclear feedback can result in an X-ray 
cavity, as hinted at by the central flattening of the X-ray intensity (Fig.~\ref{fig:surbv}c and Fig.~\ref{fig:multi}b). Another possible signature of the feedback is the energetic particles inferred
to be present along the nuclear spiral (\S~\ref{sec:obs}), although it is also possible that they originate
from SN events. High-resolution, high-sensitivity radio observations will help to clarify this issue.

\subsection{Origin, role, and fate of the hot gas} {\label{subsec:role}}
The study of the circumnuclear diffuse X-ray emission (\S~\ref{subsec:hotgas}) reveals the unambiguous
presence of hot gas in the core of M31. The next question to address is what supplies the hot gas. Diffuse hot gas is 
commonly found in the cores of early-type galaxies. Proposed origins of hot gas  
include accretion of the intergalactic medium (IGM), typically prevalent in massive, high-$L_X$ elliptical galaxies,
and the collective ejecta of local evolved stars, likely predominant in low-$L_X$ early-type galaxies. In the latter case,
the most important heating source of the stellar ejecta is thought to be Type Ia SNe (e.g., Ciotti et al.~1991; David et al.~2006). 
This should be the case in the M31 bulge, where the SMBH is quiescent and there is no recent massive star formation. 
In the present-day universe, a stellar spheroid empirically deposits energy and mass
at rates of $\sim$$1.1\times10^{40}$$[L_K/(10^{10} L_{\odot,K})]$${\rm~ergs~s^{-1}}$ (Mannucci et al.~2005) 
and $\sim$$2{\times}10^{-2}[L_K/(10^{10} L_{\odot,K})]$${\rm~M_\odot~yr^{-1}}$ (Knapp, Gunn \& Wynn-Williams 1992), respectively, 
given an energy release of 10$^{51} {\rm~ergs}$ per SN Ia. Assuming an SN heating efficiency $\epsilon\sim$1 and that the stellar mass loss is 
wholly involved, an 
average energy input of $\sim$3.6 keV per gas particle is inferred. The gravitational potential of a normal galaxy 
is unlikely to confine the gas with such a high temperature. Therefore the gas is expected to escape, at least from inner regions
of the host galaxy. Indeed, for many early-type galaxies the observed X-ray luminosity of hot gas, $L_X$, is typically no more than a few 
percent of the expected energy
input rate from SNe Ia (e.g., David et al.~2006; Li, Wang \& Hameed 2007; LW07); the inferred gas mass is also
much less than expected if the stellar ejecta has been accumulating for a substantial fraction of the host galaxy's lifetime.
Such discrepancies can be explained naturally with the presence of an outflow of hot gas, in which the ``missing'' energy and mass 
are transported outside the regions covered by the observations.

It can be shown that the stellar feedback is also largely missing in the core of M31.
For the central arcmin we have $L_X{\approx}7{\times}10^{37}{\rm~ergs~s^{-1}}$ and $M_h{\approx}10^5{\rm~M_\odot}$, 
whereas the SN heating rate is $\sim$$5{\times}10^{39}{\rm~ergs~s^{-1}}$ and the mass input rate
is $\sim$$0.01{\rm~M_\odot~yr^{-1}}$ (taking only $\sim$$10^7 {\rm~yr}$ to accumulate the observed amount of hot gas).
Clearly, these suggest an outflow launched in the circumnuclear regions, as already hinted at by the X-ray morphology (\S~\ref{subsec:hotgas}).
On the other hand, the fitted gas temperature in the M31 core, $\sim$0.3 keV (\S~\ref{subsec:hotgas}), is much lower than 
the maximum allowed temperature of $\sim$3.6 keV ($\epsilon \sim 1$), 

That $\epsilon$$\sim$1 is often implicitly assumed in the context of early-type galaxies, in which 
individual SNe occur in volume-filling hot gas.
The SN blast wave does not dissipate easily and can effectively convert mechanical energy into thermal energy (i.e., $\epsilon$$\sim$1; Tang \& Wang 2005). 
In the case of M31, dissipation of the blast wave
could become important when it encounters the cold, dense gas of the nuclear spiral.
The amount of energy loss in such a process, via mechanical and radiational dissipations, depends on the geometry of the encounter and the local gas density (e.g., McKee \& Cowie 1975), but is unlikely large
enough to result in $\epsilon < 0.5$, as the solid angle covered by the nuclear spiral with respect to the wave front must be less than 2$\pi$.

Alternatively, a reduced gas temperature is expected given a mass input in addition to the stellar deposition. In particular, this is conceivable via thermal evaporation of the nuclear spiral. Such evaporation at an appropriate rate would result 
in a ``mass-loading'' of the hot gas, a process we shall now consider. 
We assume that the nuclear spiral is composed of discrete cloudlets.
The geometry of these cloudlets, which determines the evaporation rate, is rather uncertain. As a first order approximation
we consider spherical cloudlets having a characteristic radius $R_s=R_{\rm pc} {\rm~pc}$. 
Only those cloudlets that are smaller than a critical radius, $R_{\rm crit}  
\approx 4.8\times10^5 T_h^2n_h^{-1}{\rm~cm} \approx$ 25 pc, would be evaporating; condensation would be more likely to take place for larger ones upon radiative cooling
(McKee \& Cowie 1977). While the large CO Clump may be an example of this latter
case, most cloudlets in the region, essentially unresolved under the few-arcsec resolution, appear to be small enough to undergo evaporation. 
In regions where the H$\alpha$ and MIR emission are associated,
the cold neutral gas most likely resides in the cores of the cloudlets with evaporating, ionized outer layers;
in regions where only the H$\alpha$ emission is prominent, the cloudlets are likely in the late stages of evaporation and nearly fully ionized.
The number of cold cloudlets, $N_c$, satisfies 
$(4{\pi}/3)R_s^3 N_c n_cm_H = M_c$, or $N_c R_{\rm pc}^3 \approx 960 (M_c/10^6{\rm~M_{\odot}})$.
Similarly, for the warm cloudlets we have $(4{\pi}/3)R_s^3N_w n_wm_{\rm H} = M_w$, or $N_w R_{\rm pc}^3 \approx 160$.
If $N_c \sim N_w$, the size of the warm cloudlets is about half that of the cold cloudlets,
consistent with a reduced size due to early evaporation.
The classical thermal conductivity (Spitzer 1962) holds when the mean free path of conducting
electrons, $\lambda \approx 10^4 T_{h}^2n_h^{-1} {\rm~cm} \approx 0.5 {\rm~pc}$, is shorter than 
the scale-length of temperature variation, roughly being the cloudlet radius. 
Adopting $R_{\rm pc} = 0.5$ and assuming that classical evaporation applies, the 
total evaporation rate of the cold cloudlets is $\dot{M}_{\rm evap} \approx 2.75\times10^{4}T_h^{5/2}R_{\rm pc}N_c {\rm~g~s^{-1}}
\approx 0.05{\rm~M_{\odot}~yr^{-1}}$ (Cowie \& McKee 1977), implying a corresponding mass input rate about five times that of the stellar ejecta to the $r \lesssim$ 230 pc 
region. An additional evaporation of $\sim$0.01${\rm~M_{\odot}~yr^{-1}}$ may arise from the fully ionized cloudlets. 
For cloudlets of smaller initial sizes, the conduction becomes saturated and the evaporation rate from individual cloudlets
drops significantly (Cowie \& McKee 1977), but the correspondingly number of cloudlets must increase to conserve
the total gas mass and hence could result in a comparable $\dot{M}_{\rm evap}$. 

Several self-consistency checks are needed before concluding the likelihood of the above scenario. 
First, the evaporation timescale is $t_{\rm evap}\sim2\times10^7{\rm~yr}$, not much longer than the time needed
for the cold gas to spiral an angle of 2$\pi$, 
$t_{\rm evap}\approx 2{\pi}r/v_c \sim5\times10^6{\rm~yr}$. Thus the gas is likely to have evaporated before
spiraling into the very central regions. This naturally explains the observed X-ray enhancement coincident
with the Inner Arm as well as the lack of cold gas observed in the $r \lesssim$ 50 pc regions (\S~\ref{subsec:multi}; 
Fig.~\ref{fig:multi}). Second, although difficult to measure, a gas inflow rate of $\sim$$0.05{\rm~M_{\odot}~yr^{-1}}$
from the outer disk regions of M31 is entirely possible. Typical gas inflow rates of $0.1$-$1 {\rm~M_{\odot}~yr^{-1}}$ are 
suggested for barred spirals (e.g., Quillen et al.~1995; Sakamoto et al.~1999), which often lead
to an observed concentration of $10^8$-$10^9{\rm~M_{\odot}}$ neutral gas in the central kpc. In contrast, the gas mass in the central kpc 
of M31 is merely $\sim$$10^7{\rm~M_{\odot}}$. Without consumption of gas by star formation, this lack of gas concentration is most likely due to the evaporation 
process discussed above. In this regard, 
the maximum evaporation rate afforded by a hot corona in barred spirals is likely $\sim$$0.1{\rm~M_{\odot}~yr^{-1}}$.
Third, the implied mass-loading rate to the corona results in a reduced heating per gas particle that is about right for
the measured temperature of the hot gas. The mass-loading also predicts a dilution of metallicity 
for the hot gas. In particular, the iron abundance is expected to be $Z_{\rm Fe} \approx 10 ({\rm M_{Fe}}/0.7 {\rm M_\odot})$
due to SN Ia enrichment (Nomoto, Thielemann \& Yokoi 1984) of the collective stellar ejecta alone. The measured $Z_{\rm Fe} \sim 1-2$ solar (Table \ref{tab:spec}),
is consistent with the predicted dilution. It is not immediately clear, however, how a supersolar abundance of Mg
and a subsolar abundance of O are simultaneously derived from the mass-loading.  One possibility is that the Mg abundance reflects the primary Type II SNe 
enrichment at the outer disk regions before gas inflowing to the nuclear spiral. 
The O abundance, on the other hand, might have been
underestimated, in a sense that the resonance scattering of the O VII 
and O VIII K$\alpha$ line emission is not properly accounted for in the spectral fit,
since these lines are optically thick if the velocity dispersion of the hot gas 
is not far from its thermal broadening.

\subsection{Origin of the \OVI\ absorption lines} {\label{subsec:uv}}

The detection of the \OVI\ absorption lines (\S~\ref{subsec:ovi}) provides direct evidence for the presence of interstellar gas with temperatures of a few $10^5$ K associated with M31. 
The next question to address is the origin of this gas. 
In view of the discussed scenario above, gas with intermediate temperatures (between $10^4-10^6$ K) 
is naturally expected to exist in the evaporating layers of the nuclear spiral, 
which may be enhanced by turbulent mixing (Slavin, Shull, \& Begelman 1993) induced by the SNe Ia shock waves.  
For individual classically evaporating cloudlet, the absorption column is (McKee \& Cowie 1977),
\begin{equation}
N_{\rm OVI,c} = \int 2.5(T/T_h)^{1/2} n_h X_{\rm O^{5+}}[n_{\rm O}/n_{\rm H}]R_s d(T/T_h) \approx 5.2\times10^{18} n_h X_{\rm O^{5+}}[n_{\rm O}/n_{\rm H}]R_{\rm pc} {\rm~cm^{-2}},
\end{equation}
where $X_{\rm O^{5+}}$ is the ionization fraction of ${\rm O^{5+}}$ ions. 
In collisional ionization equilibrium (CIE), $X_{\rm O^{5+}}$ peaks at $\sim$0.24 at a temperature 
of $\sim$$3\times10^5$ K (Mazzotta et al.~1998), whereas in the evaporating layers where the ionization structures
are determined by the gas dynamics and typically of non-equilibrium (NIE), $X_{\rm O^{5+}}$ can be as high as $\sim$0.6 (B${\rm \ddot{o}}$hringer \& Hartquist 1987).  
Taking the NIE value of $X_{\rm O^{5+}}$, a solar oxygen abundance $[n_{\rm O}/n_{\rm H}] \approx 6.8\times10^{-4}$ (Grevesse \& Sauval 1998), and our fiducial values of 
$n_h$ and $R_{\rm pc}$ (\S~\ref{subsec:role}), we have $N_{\rm OVI,c} \sim 10^{14}{\rm~cm^{-2}}$.
Therefore, a circumnuclear origin for the observed column of $4\times10^{14} {\rm~cm^{-2}}$ requires on average multiple
conductive interfaces along the sight-line, 
which does not seem warranted given the small volume filling factor ($f$$\sim$$10^{-4}$) of the cloudlets.
Moreover, while the conductive interfaces are likely to absorb the bulge starlight from the far-side (i.e., behind
the nuclear spiral) to some extent, the starlight from the near-side bulge likely compensates
the absorption to the extent that no prominent \OVI\ absorption lines would be detected, if \OVI-absorbing gas only exists in the 
circumnuclear regions. The fact that a column density as high as $4\times10^{14} {\rm~cm^{-2}}$ is observed thus implies
a subtantial fraction of the \OVI-absorbing gas distributed in and around the near-side bulge. 

In a two-dimensional spectroscopic study covering approximately the same field of view as the {\sl FUSE} observations, del Burgo
et al.~(2000) identified several features of ionized gas, which they desigated as clouds A-F, via H$\beta$, [O III],
H$\alpha$ and [N II] emission lines. At resolutions of 0\farcs5-2\farcs7 ($\sim$2-10 pc), these clouds are found to 
have typical diameters of $\sim$2 pc and $\gtrsim$10 pc. The small ones exhibit a narrow velocity range and are likely truly isolated 
with a size consistent with that of our fiducial cloudlets, 
whereas the larger ones tend to have an elongated
shape and exhibit velocity gradients and thus are likely composed of multiple cloudlets. The cloud velocities (corrected for the systemic velocity of $-$310 ${\rm km~s^{-1}}$) range from $-256$ to 234 ${\rm km~s^{-1}}$, 
with a mean value of $-39\pm9{\rm~km~s^{-1}}$ and a typical velocity dispersion of 60 ${\rm km~s^{-1}}$. 
Thus the \OVI-absorbing gas responsible for the observed lines with a centroid velocity
of $\sim$40 ${\rm km~s^{-1}}$, blueshifted with respect to the systemic velocity of M31 (\S~\ref{subsec:ovi}), 
appears to be dynamically distinct from the circumnuclear ionzied gas. This supports the above argument that  
a subtantial fraction of the observed \OVI\ column density arises from the near-side bulge.
Moreover, the blueshifted velocity of the \OVI-absorbing gas implies that it is
also not dynamically coupled with the global outflow of hot gas.
A probable origin of the \OVI-absorbing gas is the ejecta of bulge stars (e.g., red giant winds,
planetary nebulae) being heated by the ambient hot gas (Parriott \& Bregman 2008).
A quantitative accounting of the observed quantities demands a proper modelling of the stellar SED, the distribution
and evolution of the stellar ejecta and radiation transfer along the sightline, and is beyond the scope of the present study. Alternatively, the observed \OVI\ column density may mostly arise from in front of the bulge, i.e., the highly inclined thick disk of M31. If the amount of \OVI-absorbing gas in the M31 disk is similar to that of our Galaxy, we
would then expect an amount of $10^{14.2}{\rm cm^{-2}}/{\rm sin}(12.5 {\rm~deg}) \approx 7.3\times10^{14} {\rm~cm^{-2}}$, where 12.5 deg accounts for the inclination of the disk and $10^{14.2}{\rm~cm^{-2}}$ is the average
vertical column density observed in the Galaxy (Savage et al.~2000). Of course, the \OVI\ column varies by a factor of 2 between the two sides of
the Galactic disk, and the amount of \OVI-absorbing gas is probably less in M31 than in the Galaxy,
as the latter is substantially more active in star formation. Nevertheless, within the
uncertainties, the bulk of the observed \OVI\ column density can be attributed to the highly inclined thick disk of M31.

\begin{figure}[!htb]
  \centerline{
          \epsfig{figure=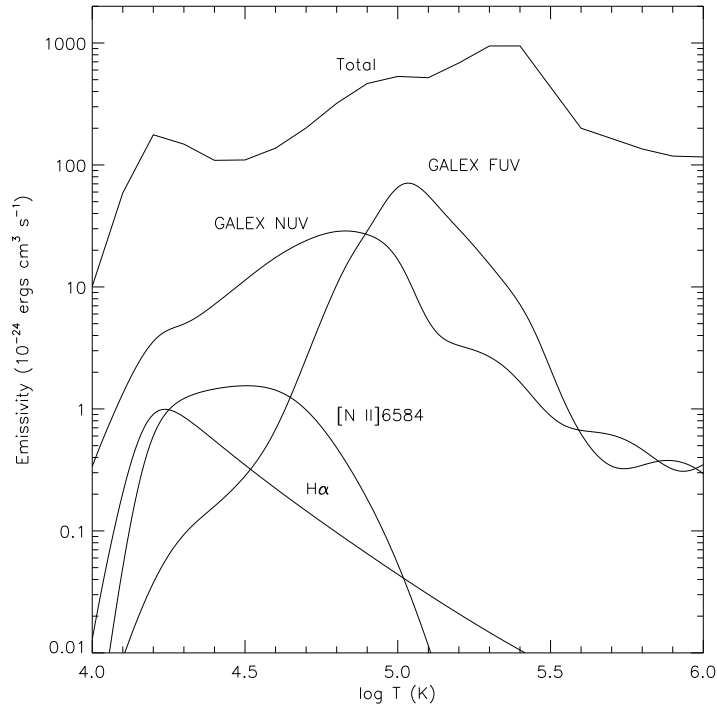,width=0.6\textwidth,angle=0}
          }
	      \caption{Temperature-dependent volume emissivities of selected
lines and bands, for a gas of solar abundance in CIE.}
\label{fig:uv}
\end{figure}

In Fig.~\ref{fig:multi}d we have noted the FUV enhancement apparently associated with 
the nuclear spiral. It is possible that this enhancement arises from the evaporating gas with a 
temperature of few $10^5$ K. 
This is illustrated in Fig.~\ref{fig:uv}, in which CIE volume emissivities of the {\sl GALEX} FUV and NUV bands are shown as a function of gas temperature. While NIE is more likely the case for the evaporating gas, 
we expect reasonable order-of-magnitude estimates based on CIE values.
At a temperature of $\sim$$10^5$ K, gas exhibits its peak FUV emissivity and an FUV/NUV ratio of $\sim$5. 
Ideally, one wants to isolate and quantify the FUV emission intrinsic to the evaporating gas, but this is hampered by the large uncertainties in the stellar SED and dust extinction. 
We derive a not quite tight constraint for the \OVI\ absorption column density, $\leq$$10^{16}{\rm~cm^{-2}}$, associated with the evaporating gas,  
from the total FUV flux of $\sim$$6.2\times10^{-14}{\rm~ergs~s^{-1}}$ observed within the central 15$^{\prime\prime}$.

\subsection{Ionizing sources of the nuclear spiral} {\label{subsec:ionize}}
With the above findings and scenarios in mind, we now revisit the ionizing source of the nuclear spiral in M31, a long-standing puzzle (Devereux et al.~1994). 
Assuming a [N II]/H$\alpha$ intensity ratio of 2 and that every recombination produces
0.45 H$\alpha$ photons, the observed H$\alpha$+[N II] flux of $1.1\times10^{-11} {\rm~ergs~s^{-1}~cm^{-2}}$ within the 
central arcmin implies a recombination rate $R \approx 2\times10^{50} {\rm~s^{-1}}$.
In the absence of massive young stars, 
possible sources for the ionizing photons include: i) UV radiation
of old stars; ii) X-ray photons from the hot gas as well as stellar objects; iii) UV photons induced by thermal conduction; 
iv) UV photons induced by shocks;
and v) cosmic-rays produced by the nucleus or SN events. Below we assess the relative importance of these sources.

Extended recombination line emission, particularly H$\alpha$ emission, is often observed in elliptical galaxies. 
Binette et al.~(1994) demonstrated that the ionizing photons from post-AGB stars with effective temperatures of $\sim$10$^5$ K 
are typically sufficient to account for the observed H$\alpha$ emission. 
The H$\alpha$ intensity is found to be correlated with the optical luminosity within regions
of the line emission (Macchetto et al.~1996), further suggesting a stellar origin of the ionizing photons in elliptical galaxies. 
According to Binette et al.~(1994), post-AGB stars provide ionizing photons (i.e., shortward of 912${\rm~\AA}$)
at a rate of $7.3\times10^{40} {\rm~s^{-1}~M_\odot^{-1}}$. In the central arcmin of M31, 
this gives a recombination rate of $R \approx 2.9g\times10^{50} {\rm~s^{-1}}$, where $g \lesssim 0.5$ is a geometric factor
that determines the fraction of ionizing photons received by the nuclear spiral. 

It is also possible that the ionizing photons are related to the so-called {\sl UV-upturn}, a rise 
in the SED shortward of $\sim$2500${\rm~\AA}$ observed in elliptical galaxies 
as well as the M31 core (Burstein et al.~1988). 
In view of the lack of young stars, the UV-upturn is generally attributed to the emission of hot horizontal branch (HB) stars 
(e.g., Brown et al.~1998), with an effective temperature of a few $10^4$ K. Such stars could also be partially responsible 
for the recombination line emission observed in M31. According to the latest model for the UV-upturn (Han et al.~2007; Han 2008, 
private communication), HB stars produce ionizing photons at a rate of $1.4\times10^{40} {\rm~s^{-1}~M_\odot^{-1}}$,
corresponding to a recombination rate of $R \approx 5.6g\times10^{49} {\rm~s^{-1}}$ in the central arcmin of M31.

Ionization could also be induced by X-rays via secondary ionization by photoelectrons. 
Assuming that on average each incident X-ray photon from the hot gas produces a 0.3 keV primary photoelectron,
the number of secondary ionizations is $\sim$10 (Shull 1979). The recombination rate induced by X-ray photons 
is then $R \approx 1.2g\times10^{48} {\rm~s^{-1}}$, according to the photon luminosity of $\sim$$1.2\times10^{47}{\rm~s^{-1}}$ within the central arcmin. While
stellar objects dominate the overall X-ray luminosity of M31,
there are only 33 sources detected within the central arcmin
(Voss \& Gilfanov 2007),
the corresponding covering factor $g$ of which must be small compared to that of the volume-filling hot gas.
Hence the X-ray ionization by stellar objects is negligible. 
   
The thermal conduction proposed above is accompanied by ionizations. Ionizing photons are generated at a rate of $R \approx 2\times10^{-8}n_h^2T_h^{-0.6}R_sS{\rm~s^{-1}}$ (McKee \& Cowie 1977), 
where $S = 4{\pi}R_s^2N_c$ is the effective area of the conduction front and other symbols are defined above. 
With $N_cR_{pc}^3 = 960$, $R \approx 8\times10^{45} {\rm~s^{-1}}$. 

Shocks are likely present in the nuclear spiral either due to orbital dissipation (e.g., Englmaier \& Shlosman 2000) or the propagation
of SN blast wave into the cold gas. Such shocks produce ionizing photons at a rate of $R = {\kappa}n_cv_sS$ (Shull \& McKee 1979),
where $v_s$ is the shock velocity, ${\kappa}$ a tabulated numerical factor and $S$ the rather uncentain effective area of the shock front. 
Adopting $v_s = 40 {\rm~km~s^{-1}}$ and $S = (100{\rm~pc})^2$, we have $R \approx 3\times10^{48} {\rm~s^{-1}}$. 
  
Finally, ionizations can be induced by cosmic-rays. 
We adopt an ionization rate per hydrogen atom of $5.4f^{-4/7}\times10^{-15} {\rm~s^{-1}}$ for 2 MeV protons (Goldsmith, Habing \& Field 1969), with an energy density of $0.5f^{-4/7} {\rm~eV~cm^{-3}}$ inferred from the central regions of M31 (\S~\ref{sec:obs}).
With a hydrogen mass of $10^3{\rm~M_\odot}$, the recombination rate is inferred to be $R \approx 1.2\times10^{48}{\rm~s^{-1}}$, given a filling factor $f \approx 10^{-4}$.
  
From the above estimates, it is evident that the only likely ionizing source for the nuclear spiral is the stellar UV radiation predominantly
contributed by the post-AGB stars, with an additional contribution from HB stars. The other sources all fall short of accounting for the required ionizing photons. 
The stellar radiation being the ionizing source is consistent with the
observed limb-brightened H$\alpha$ emission of the filaments on the sides
facing the M31 center (\S~\ref{subsec:multi}), where the radiation 
intensity presumably peaks. 
However, it has been a concern that models for the photoionization due to post-AGB stars predict an [N II]$\lambda$6584/H$\alpha$ intensity ratio of $\sim$1.2 
for gas with an abundance up to 3 solar (Binette et al.~1994), which is 
inconsistent with the high ratios (generally $\gtrsim$1.3, as large as $\sim$2.7) observed in M31 (Ciardullo et al.~1988). 
The relatively high [N II]/H$\alpha$ ratio implies that heating in addition to photoionization contributes substantially to the production of the nitrogen ions.
For instance, a temperature of $\sim$$10^{4.35}$ K is needed for gas in CIE to produce a [N II]/H$\alpha$ ratio of 2 (Fig.~\ref{fig:uv}), while at temperatures below $10^{4.2}$ K the ratio is essentially below 1.

If a substantial fraction of the observed [N II] line emission indeed arises from gas heated to $\sim$$10^{4.5}$ K, 
the total [N II] luminosity of $\sim$$5\times10^{38}{\rm~ergs~s^{-1}}$ then implies a heating rate on the order 
of $10^{40}{\rm~ergs~s^{-1}}$, as the [N II] lines typically account for only few percent of the total radiative cooling (Fig.~\ref{fig:uv}).
The energy input from SNe Ia, while responsible for the heating of the hot gas, alone may not be sufficient to 
account for this radiative loss. 
In this regard, energetic particles ejected by the nucleus may serve as an additional heating source, given the above estimated jet power of $1.6\times10^{40}{\rm~ergs~s^{-1}}$.

\section{Summary and concluding remarks} {\label{sec:sum}}
We have carried out a comprehensive analysis of the multiwavelength data on the circumnuclear environment of M31. We summarize our major results and comment on relevant prospects as follows:    
\begin{itemize}
\item We have derived a tight constraint on the X-ray luminosity of M31$^\ast$, $L_{0.3-7 {\rm~keV}} \lesssim 1.2\times10^{36}{\rm~ergs~s^{-1}}$. 
The estimated jet power from M31$^\ast$, $1.6\times10^{40}{\rm~ergs~s^{-1}}$, probably significantly contributes to balancing the radiative cooling of the nuclear spiral.
Future high-sensitivity, long-duration and simulteneous X-ray/radio observations may lead to the
detection of timing variabilities intrinsic to M31$^\ast$ and help to establish 
its appearance in X-ray as suggested by the present study. Such 
observations are also crucial for assessing the relative importance of feedback from inactive nuclei. 

\item We have determined a temperature of 0.3 keV and a mass of $\sim$$10^5 {\rm~M_{\odot}}$ for the circumnuclear X-ray-emitting hot gas within the central 230 pc. For the dusty cold gas residing in
the 
nuclear spiral, we have revealed its dust emission and morphological evidence of its interaction with the hot gas.

\item We have proposed a self-consistent scenario for understanding much of the multiwavelength phenomena,
in which thermal conduction between the hot corona and the nuclear spiral plays a crucial role
in regulating the evolution of the SMBH and the circumnuclear ISM.
Further tests of the scenario are needed, including high-resolution imaging-spectroscopic observations of 
optical and UV emission lines arising from the conductive interfaces.
The scenario, albeit crude in details at the moment, should have important applications for 
similar circumnuclear environments of spiral galaxies, in particular our Galaxy.
It is also reasonable to invite its further application to elliptical galaxies (e.g., Sparks, Macchetto \& Golombek 1989), 
in which hot and cold gas are often observed to co-exist and sometimes show morphological 
correlations (e.g., Trinchieri, Noris \& di Serego Alighieri 1997). In the absence of a large-scale disk,
elliptical galaxies may obtain a substantial supply of cold gas from galaxy mergers. In any case,
thermal conduction is expected to be prevalent in the core of early-type galaxies containing typically
dense, multi-phase ISM, a process previously overlooked. Numerical modeling of such a scenario will 
help to derive characteristics of the dynamical and thermal properties of both the hot and cold gas.

\item We have detected \OVI\ absorption against the bulge UV starlight. While the circumnuclear 
conductive interfaces may give rise to \OVI\ absorption, a subtantial fraction of the observed absorption
column density probably arises from the outer bulge and/or the highly inclined disk. 

\item Understanding the ISM evolution in the circumnuclear regions also allows for a better understanding
of its evolution on large scales. A detailed modeling of the hot gas as well as the \OVI-absorbing gas 
in the M31 bulge will be the goal of a forthcoming paper.

\end{itemize}
\vskip 0.5cm
We are indebted to Dr.~N.~Devereux for providing the H$\alpha$ image and to Dr.~L.~Ji for providing the numerical data
ultilized in Fig.~\ref{fig:uv}. It is a pleasure to thank Drs.~D.~Calzatti, M.~Garcia, Z.~Han and L.~Ho for 
stimulating and helpful discussions. This work is supported by the SAO 
grant AR7-8006X.

\end{document}